\documentclass[a4paper,fleqn,usenatbib,useAMS]{mnras}

\usepackage{graphicx}	
\usepackage{amsmath}	
\usepackage{amssymb}	
\usepackage{multicol}        
\usepackage{bm}		
\usepackage{pdflscape}	

\usepackage{mathtools}

\usepackage[shortlabels]{enumitem}

\defcitealias{McClure-Griffiths_2010}{McG10}

\newcommand{\lsim}{{\;\raise0.3ex\hbox{$<$\kern-0.75em\raise-1.1ex\hbox{$\sim$}}\;}}
\newcommand{\gsim}{{\;\raise0.3ex\hbox{$>$\kern-0.75em\raise-1.1ex\hbox{$\sim$}}\;}}

\usepackage[T1]{fontenc}
\usepackage{ae,aecompl}

\usepackage{newtxtext,newtxmath}

\title[The Faraday rotation sky of TNG50]{Sampling the Faraday rotation sky of TNG50: Imprint of the magnetised circumgalactic medium around Milky Way-like galaxies}

\author[Jung et al.]{S. Lyla Jung$^{1}$\thanks{e-mail: \href{mailto:lyla.jung@anu.edu.au}{lyla.jung@anu.edu.au}}, N. M. McClure-Griffiths$^{1}$, Rüdiger Pakmor$^{2}$, Yik Ki Ma$^{1}$, Alex S. Hill$^{3}$, 
\newauthor  Cameron L. Van Eck$^{1}$, Craig S. Anderson$^{1}$
\\
\\
$^{1}$ Research School of Astronomy \& Astrophysics, The Australian National University, Canberra ACT 2611, Australia
\\
$^{2}$ Max Planck Institute for Astrophysics, Karl-Schwarzschild-Str. 1, 85741 Garching, Germany
\\
$^{3}$ Department of Computer Science, Math, Physics, and Statistics, the University of British Columbia, Okanagan Campus, 3187 University Way, Kelowna,\\ BC V1V1V7, Canada
}

\date{Last updated}

\pubyear{2023}

\begin{document}
\label{firstpage}
\pagerange{\pageref{firstpage}--\pageref{lastpage}}
\maketitle

\begin{abstract}
Faraday rotation measure (RM) is arguably the most practical observational tracer of magnetic fields in the diffuse circumgalactic medium (CGM). We sample synthetic Faraday rotation skies of Milky Way-like galaxies in TNG50 of the IllustrisTNG project by placing an observer inside the galaxies at a solar circle-like position. Our synthetic RM grids emulate specifications of current and upcoming surveys; the NRAO VLA Sky Survey (NVSS), the Polarisation Sky Survey of the Universe's Magnetism (POSSUM), and a future Square Kilometre Array (SKA1-mid) polarisation survey. It has been suggested that magnetic fields regulate the survival of high-velocity clouds. However, there is only a small number of observational detections of magnetised clouds thus far. In the first part of the paper, we test conditions for the detection of magnetised circumgalactic clouds. Based on the synthetic RM samplings of clouds in the simulations, we predict upcoming polarimetric surveys will open opportunities for the detection of even low-mass and distant clouds. In the second part of the paper, we investigate the imprint of the CGM in the all-sky RM distribution. We test whether the RM variation produced by the CGM is correlated with global galaxy properties, such as distance to a satellite, specific star formation rate, neutral hydrogen covering fraction, and accretion rate to the supermassive black hole. We argue that the observed fluctuation in the RM measurements on scales less than $1^{\circ}$, which has been considered an indication of intergalactic magnetic fields, might in fact incorporate a significant contribution of the Milky Way CGM.
\end{abstract}





\section{Introduction}
Galaxies are surrounded by the circumgalactic medium (CGM) that evolves interactively with the interstellar medium (ISM) and the intergalactic medium (IGM) beyond the galactic halo. 
While most of the volume within the halo is filled with hot diffuse gas, observations report the detection of denser and cooler phase gas clouds around the Milky Way as well as nearby galaxies (e.g., \citealt{Westmeier_2018}; \citealt{Lehner_2020}). Observationally, such circumgalactic gas clouds around the Milky Way are referred to as high- or intermediate-velocity clouds (HVCs or IVCs; see \citealt{Putman_2012} and \citealt{Richter_2017} for a review).
HVCs and IVCs are associated with diverse thermodynamical mechanisms in the CGM, for example, cold accretion of cosmic filaments, cooling of energetic outflows from the Galactic disk, and ram pressure/tidal stripping of cold gas in satellite galaxies (\citealt{Wakker_1997}; \citealt{Westmeier_2007}; \citealt{Oosterloo_2007}; \citealt{Olano_2008}; \citealt{Putman_2012}; \citealt{Fraternali_2015}; \citealt{Marasco_2017}).

The magnetic field strength in the galactic halo is usually very weak ($\sim0.1\rm \mu G$, \citealt{Han_1994}; \citealt{Taylor_2009}; \citealt{Jansson_2012}), but fast-moving clouds can sweep up and stretch the ambient halo magnetic field lines and amplify the field strength around them.
Magneto-hydrodynamic (MHD) numerical simulations suggest that magnetic fields draped and amplified around a cloud regulate the mixing of gas at the cloud-halo interface, thus affecting the amount of radiative cooling in the cloud system, and eventually helping the survival of the cloud throughout the passage within the halo (\citealt{Jones_1996}; \citealt{Gregori_1999, Gregori_2000}; \citealt{Santillan_1999}; \citealt{Dursi_2008}; \citealt{Shin_2008}; \citealt{Kwak_2009}; \citealt{McCourt_2015}; \citealt{Banda-Barragan_2016, Banda-Barragan_2018}; \citealt{Gronnow_2017, Gronnow_2022}; \citealt{Gronke_2020}; \citealt{Cottle_2020}; \citealt{Sparre_2020}; \citealt{Jung_2023}; \citealt{Bruggen_2023}; \citealt{Hidalgo-Pineda_2023}).

Faraday rotation of background polarisation sources (e.g., quasars and pulsars) has been a major observational tracer for searching for magnetised circumgalactic clouds around the Milky Way. The rotation measure (RM) is a measure of the change in the polarisation angle due to Faraday rotation as linearly polarised radiation propagates within a magneto-ionic medium:
\begin{equation}\label{eq:RM}
    RM = 0.812\int_{\rm observer}^{\rm source}n_{\rm e}(r)B_{\parallel}(r)dr,
\end{equation}
where $RM$ is in units of $\rm rad\,m^{-2}$, $n_{\rm e}$ is the electron density in $\rm cm^{-3}$, $B_{\parallel}$ is the magnetic field strength along the line-of-sight in $\mu G$, and $r$ is a path length in $\rm pc$. 
There have been reports of three high-velocity HI complexes around the Milky Way that spatially overlap with the excessive RM in the Faraday rotation sky: the Magellanic Leading Arm (\citealt{McClure-Griffiths_2010}); the Smith cloud (\citealt{Hill_2013}; \citealt{Betti_2019}), and the Magellanic bridge (\citealt{Kaczmarek_2017}). 
Although, the RM excess in the Magellanic Leading Arm region has recently been demonstrated to be a contribution from an overlapping object (\citealt{Jung_2021}).
There are more detections when extending the search to IVCs or ionized clouds (e.g., \citealt{Stil_2016}).
Still, little is known about the magnetic field properties of the vast majority of the hundreds of circumgalactic clouds.

There have been attempts to study magnetised CGM around external galaxies using RM of polarised sources (e.g., \citealt{Lan_2020, Heesen_2023, Bockmann_2023}). 
Albeit not resolved to scales as small as the Milky Way's CGM, challenges are in part mitigated by stacking sources in the background of an assortment of external galaxies. Such stacking experiments are suitable for obtaining global profiles of magnetic properties in the CGM, rather than focusing on magnetic fields associated with individual circumgalactic structures.
For example, \citet{Heesen_2023} report enhanced RM values within impact parameters of $100\,\rm kpc$ of nearby galaxies specifically along the minor axis.
This angular dependency of the RM excess is qualitatively reproduced in numerical simulations (\citealt{Ramesh_2023c}).

This work is the first paper of a series exploring the Faraday rotation of galactic structures imprinted to the all-sky RM distribution using cosmological simulations TNG50. 
One goal of this study is to examine the detection statistics of magnetised HVCs around simulated Milky Way-like galaxies. To do so, we perform synthetic Faraday rotation measure observations of the high-resolution cosmological suite of simulations TNG50 (\citealt{Nelson_2019}; \citealt{Pillepich_2019}). 
We sample the RM distribution at different source densities and RM precision. By doing so, we investigate conditions for the detection of magnetised circumgalactic clouds and evaluate the extent to which current and future radio polarimetric surveys are capable of detecting these clouds given their observational specifications.

The second main goal of our study is to examine the imprint of the CGM as a whole (not limiting ourselves to HI clouds) onto the Faraday rotation sky of Milky Way-like galaxies in TNG50. 
The study of intergalactic magnetic fields in the large-scale structure of the Universe is a field that is expecting significant advances with future polarimetric surveys and instruments (\citealt{Heald_2020}).
Quantifying the contribution of the CGM to the all-sky RM distribution is an important piece of information that can assist with the detection of extragalactic magnetic fields. 
This is because the Faraday rotation at the Milky Way CGM inevitably adds to the net Faraday rotation of any extragalactic polarised radiation coming towards the observer. There have been suggestions from theoretical studies for methods to separate galactic and extragalactic Faraday rotation, e.g., the preferred angular scale of the imprint of Faraday rotation at the cosmic large-scale structure (\citealt{Akahori_2011}). 


We make clear that the motivation of this paper is not in testing the power of the cosmological simulations in reproducing observed properties of the CGM.
There are a number of earlier studies demonstrating that the CGM properties of Milky Way-like galaxies in current state-of-the-art cosmological simulations are generally comparable to observations (e.g., \citealt{vandeVoort_2019}).
We rather refer interested readers to related publications exploring the physical nature of cold circumgalactic clouds in TNG50, for example, \citet{Nelson_2020}, \citet{Ramesh_2023a, Ramesh_2023b}.
In this paper, we focus on utilizing the existing MHD cosmological simulations to assess the potential observability of CGM signatures under the assumption that magnetised clouds in the simulations are similar to those observed in the Milky Way.

This paper is structured as follows. We provide an overview of the TNG50 simulation in Section \ref{sec:simul} and define the Milky-Way-like galaxy sample in Section \ref{sec:MW_selection}. We describe how we define the CGM in simulations and identify individual circumgalactic clouds in Sections \ref{sec:ism_vs_cgm} and \ref{sec:cloud_id}.
In Section \ref{sec:result1}, we evaluate the capabilities of current and future radio polarimetric surveys in detecting magnetised HI circumgalactic clouds based on our synthetic Faraday rotation samplings.
In Section \ref{sec:result2}, we quantify the contribution of the entire CGM to the all-sky RM distribution and compare the results of simulated galaxies to the Milky Way observations. 
Section \ref{sec:summary} is the summary and conclusion.
We discuss the effect of the simulation resolution in Appendix \ref{sec:resol}.

\section{Method}\label{sec:method}
\subsection{A brief overview of the IllustrisTNG project} \label{sec:simul}

The IllustrisTNG project (Illustris The Next Generation, \citealt{Marinacci_2018}; \citealt{Naiman_2018};
\citealt{Nelson_2018}; \citealt{Pillepich_2018}; \citealt{Springel_2018}) is a suite of MHD cosmological simulations using the {\sc AREPO} (\citealt{Springel_2010}; \citealt{Pakmor_2016}). The simulations assume a \citet{Planck_2016} cosmology ($\Omega_{\rm \Lambda,0}= 0.6911$, $\Omega_{\rm m, 0}= 0.3089$, $\Omega_{\rm b, 0}=0.0486$, $\sigma_{\rm 8}= 0.8159$, $n_{\rm s} = 0.9667$ , and $h = 0.6774$).

In this study, we focus on TNG50 which achieves the best spatial and mass resolution among the IllustrisTNG suite.
Its relatively small simulation volume ($\sim 50^{3}\,\rm Mpc^{3}$ co-moving box) compared to TNG100 ($\sim 100^{3}\,\rm Mpc^{3}$) and TNG300 ($\sim 300^{3}\,\rm Mpc^{3}$) is not a major hurdle for our study. We have ensured a sufficient number of samples, i.e., Milky Way-like galaxies and circumgalactic clouds, as will be presented in the following sections. 
Detailed descriptions of the simulations appear in many earlier publications referenced in this paper. Specifically, we refer interested readers to the TNG50 introduction papers for details about the simulations (\citealt{Nelson_2019}; \citealt{Pillepich_2019}). 
A comprehensive description of sub-grid model treatments (e.g., star formation, chemical evolution, radiative cooling, stellar/black hole feedback) and numerical methods for the simulations are provided in the TNG methods papers (\citealt{Weinberger_2017}; \citealt{Pillepich_2018b}).
Here, we provide a brief introduction to the TNG50 suite.

\begin{table*}
\centering
\caption{The properties of the simulations used in this study. TNG50-1 is the standard resolution fiducial simulation and TNG50-2 is its lower-resolution counterpart. We present the simulation volume, the box side-length ($L_{\rm box}$), the initial number of dark matter particles ($N_{\rm DM}$) and gas cells ($N_{\rm gas}$), the mean dark matter and baryon mass resolution ($m_{\rm DM}$ and $m_{\rm gas}$), the softening of the collisionless particles at $z=0$ ($\epsilon_{\rm DM, \star}$), and the minimum allowed Plummer equivalent gravitational force softening length for gas cells ($\epsilon_{\rm gas, min}$). The last two columns show the number of Milky Way-like galaxies and the number of circumgalactic clouds identified in each simulation. See the text for the definitions.
}\label{tab:simul}
\begin{tabular}{ccccccccccccc}
\hline
&Volume& $L_{\rm box}$& $N_{\rm DM}$& $N_{\rm gas}$& $m_{\rm DM}$& $m_{\rm gas}$& $\epsilon_{\rm DM, \star}$&$\epsilon_{\rm gas, min}$& N(galaxy)& N(cloud)\\
&[Mpc$^{3}$]&[$\rm cMpc/h$]&-&-&[M$_{\odot}$]&[M$_{\odot}$]&[pc]&[pc]&-&-
\\\hline\hline
 TNG50-1 &$51.7^{3}$& 35& $2160^{3}$& $2160^{3}$& $4.5\times10^{5}$& $8.5\times10^{4}$& 288& 74& 56& 5218\\ \hline
 TNG50-2 & $51.7^{3}$& 35& $1080^{3}$& $1080^{3}$& $3.6\times10^{6}$& $6.8\times10^{5}$& 576& 148& 66& 2052\\ \hline
\end{tabular}
\end{table*}

Table \ref{tab:simul} summarises the two simulations we utilize in this paper. 
In short, TNG50-1 is a fiducial simulation and TNG50-2 is its lower-resolution counterpart that we use for a resolution test in Appendix \ref{sec:resol}.
Other than the resolution settings (the initial number of dark matter particles $N_{\rm DM}$ and gas cells, $N_{\rm gas}$; the mean dark matter mass resolution, $m_{\rm DM}$, and that of baryon, $m_{\rm gas}$; the softening length of collisionless particles, $\epsilon_{\rm DM, \star}$; and the minimum gravitational force softening length for gas cells, $\epsilon_{\rm gas, min}$), all other input parameters are identical between the two simulations.
The nature of the moving-mesh code {\sc AREPO} refines denser structures with a larger number of smaller cells.
Therefore, cold circumgalactic clouds are well resolved with $\sim 100-200 \,\rm pc$ resolution in TNG50-1 according to investigations by \citet{Nelson_2020}.

Magnetic fields evolve self-consistently in TNG50.
The simulations initially start from a homogeneous seed magnetic field with the field strength of $10^{-14}\,\rm G$ (comoving unit) at $z=127$. {\sc AREPO} advances the magnetic fields by numerically solving the ideal MHD equations (see \citealt{Pakmor_2011, Pakmor_2016, Pakmor_2013}). The divergence of magnetic fields ($\nabla \cdot \boldsymbol{B}$) is controlled using a divergence-cleaning algorithm by \citet{Powell_1999}. 
Earlier studies have shown that the magnetic field properties at low redshifts are insensitive to the seed field (\citealt{Pakmor_2014, Pakmor_2017, Marinacci_2015, Marinacci_2016, Garaldi_2021}).

\subsection{Milky Way-like sample selection}\label{sec:MW_selection}
Individual halos in the simulations are identified from the distribution of dark matter particles using the friends-of-friends (FoF) algorithm. Gravitationally bound substructures within the FoF halos, i.e., galaxies and/or subhalos, are identified using the {\sc Subfind} algorithm (\citealt{Springel_2001}).
This study focuses on the Milky Way analogy that we define based on the halo mass ($M_{\rm 200}$) and the star formation rate: $5\times10^{11}<M_{\rm 200}/M_{\odot}<3\times10^{12}$ and $0.5<SFR/(\rm M_{\odot}yr^{-1})<1.5$. 
The star formation rate is measured within $30\,\rm kpc$ from the centre of the halo and based on the total mass of stars formed in the last $250\,\rm Myr$.
In addition to the halo mass and the star formation rate, we adopt an additional criterion based on the kinematic disk-to-total ratio $f_{\rm rot}>0.7$, where $f_{\rm rot}$ is the mass fraction of the rotating gas component to the total gas in a galaxy. The exact definition of the rotating gas and the characteristic size of galaxies will be presented shortly.
Finally, we exclude merging galaxies that have substructures that are (i) within $30\,\rm kpc$ from the centre of the halo and (ii) more massive than $10\%$ of the stellar mass of the central galaxy.
There are 56 halos in the TNG50-1 simulation volume at $z=0$ that match all the above criteria.

\subsection{Separation between the galactic ISM and the CGM}\label{sec:ism_vs_cgm}

In this work, we utilize any gas within $300\,\rm kpc$ from the centre of a halo. This boundary is slightly larger than the virial radius of halos in our sample ($169<R_{\rm 200}/\rm kpc<298$).
In order to focus on the CGM of the simulated galaxies, we separate the galactic disk of the host galaxies and their CGM. We do not distinguish other substructures within the halo such as satellite galaxies and their own CGM. Instead, we consider them as the collective CGM of the host galaxy.

We use both the spatial and kinematic distributions of gas cells in the simulations to separate the ISM and the CGM.
First, we calculate the orbital circularity parameter of each gas element defined as follows (\citealt{Abadi_2003}):
\begin{equation}
    \epsilon_{\rm J}=J_{\rm z}/J_{\rm circ}(E),
\end{equation}
where $J_{\rm z}$ is the specific angular momentum of a gas element along the net spin-axis of a galaxy and $J_{\rm circ}(E)$ is the specific angular momentum of the gas if it was orbiting in a circular orbit when the orbital energy ($E$) is fixed.
The rotational axis of a galaxy is identified by calculating the net angular momentum vector of young stars (age $<1\,\rm Gyr$) within $0.01\,\rm R_{\rm 200}$.
Gas cells that follow the net rotation of a galaxy by definition have $\epsilon_{\rm J}$ close to 1.
In this work, we define gas cells with $\epsilon_{\rm J}>0.7$ as the rotating component. 

Along with the filter based on the orbital circularity parameter, we impose spatial filtering to take care of gas elements in the CGM that happen to have their rotational axis aligned with the bulk rotation of the galactic ISM. 
We determine the radius ($r_{\rm disk}$) where the mean density of the rotating gas ($\epsilon_{\rm J}$>0.7) drops below $1\%$ of the mean density within the central $10\,\rm kpc$ of the galaxy.

In brief summary, we define the rotating galactic ISM as gas cells that have (i) the orbital circularity $\epsilon_{\rm J}>0.7$ \textit{and} (ii) the distance from the galactic centre smaller than $r_{\rm disk}$.
We leave out gas cells that meet these criteria in our analysis of the CGM. 
Note that our sample of circumgalactic clouds includes the ISM of satellite galaxies as well as clouds within the halos of satellite galaxies, some of which are possibly brought into the host halo system with the infall of satellites. 
 
\subsection{All-sky projection and cloud identification}\label{sec:cloud_id}

To perform synthetic observations of the simulated galaxies, we place a mock observer at a location on the galactic mid-plane at random azimuth $8\,\rm kpc$ away from the galactic centre (i.e., the solar radius).
Then we transform the 3D coordinates of gas cells in the simulation domain from the Cartesian coordinate $(x, y, z)$ to the galactic coordinate of the mock observer $(l, b, d)$, where $d$ is the distance from the observer and galactic coordinates are defined the same as the conventional Milky Way coordinates: the galactic longitude ($l$) varies between $0$ and $360\,\rm ^{\circ}$ where the galactic centre is at $l=0\,\rm ^{\circ}$ and the galactic latitude ($b$) varies between $-90$ and $90\,\rm ^{\circ}$ 
with the galactic disk mid-plane at $b=0^{\circ}$.
The reprojected $(l, b, d)$ grid is a uniform grid as opposed to the Voronoi tessellating polyhedrons partitioning the simulation volume. 
The angular resolution of the grid is $\Delta l=\Delta b = 20\,\rm arcmin$ and for the line of sight integrals we sample along each sightline with a spatial resolution of $\Delta d = 100\,\rm pc$ unless a denser sampling of the sky is explicitly necessary or specified. In such cases, we use $\Delta l=\Delta b = 6\,\rm arcmin$ and $\Delta d = 50\,\rm pc$.
At a typical distance to the circumgalactic clouds identified in the simulation ($50\,\rm kpc$), the angular resolution of the fiducial grid corresponds to $\sim 300\,\rm pc$ physical size which is comparable to the simulation's spatial resolution for cold circumgalactic clouds.

For computing the RM along a sightline, we follow the same method described in \citet{Pakmor_2018}. 
The thermal electron density ($n_{\rm e}$) of gas cells in the simulations is calculated differently in star-forming gas and non-star-forming gas as explained in their paper: 
for non-star-forming gas cells, we use the temperature of the gas in the simulations and for star-forming gas cells, we assume a multi-phase subgrid model containing cold clumps and a warm medium (\citealt{Springel_2003}).

In this paper, especially in Section \ref{sec:result1}, we frequently refer to individual HI circumgalactic cloud complexes and discuss their properties.
For identifying these HI overdensities in the CGM, we use the friends-of-friends algorithm. 
For any cells in the $(l, b, d)$ grid with the HI column density $n_{\rm HI}>3\times10^{16}\,\rm cm^{-2}$ ($\approx 10^{-4}\,\rm cm^{-3}$ physical density given the $100\,\rm pc$ grid), we group them based on the FoF threshold separation of 3 cells ($=1^{\circ}$ in the $l$- and $b$-space and $300\,\rm pc$ in the $d$-space).  

\begin{figure*}
    \centering
    \includegraphics[width=\textwidth]{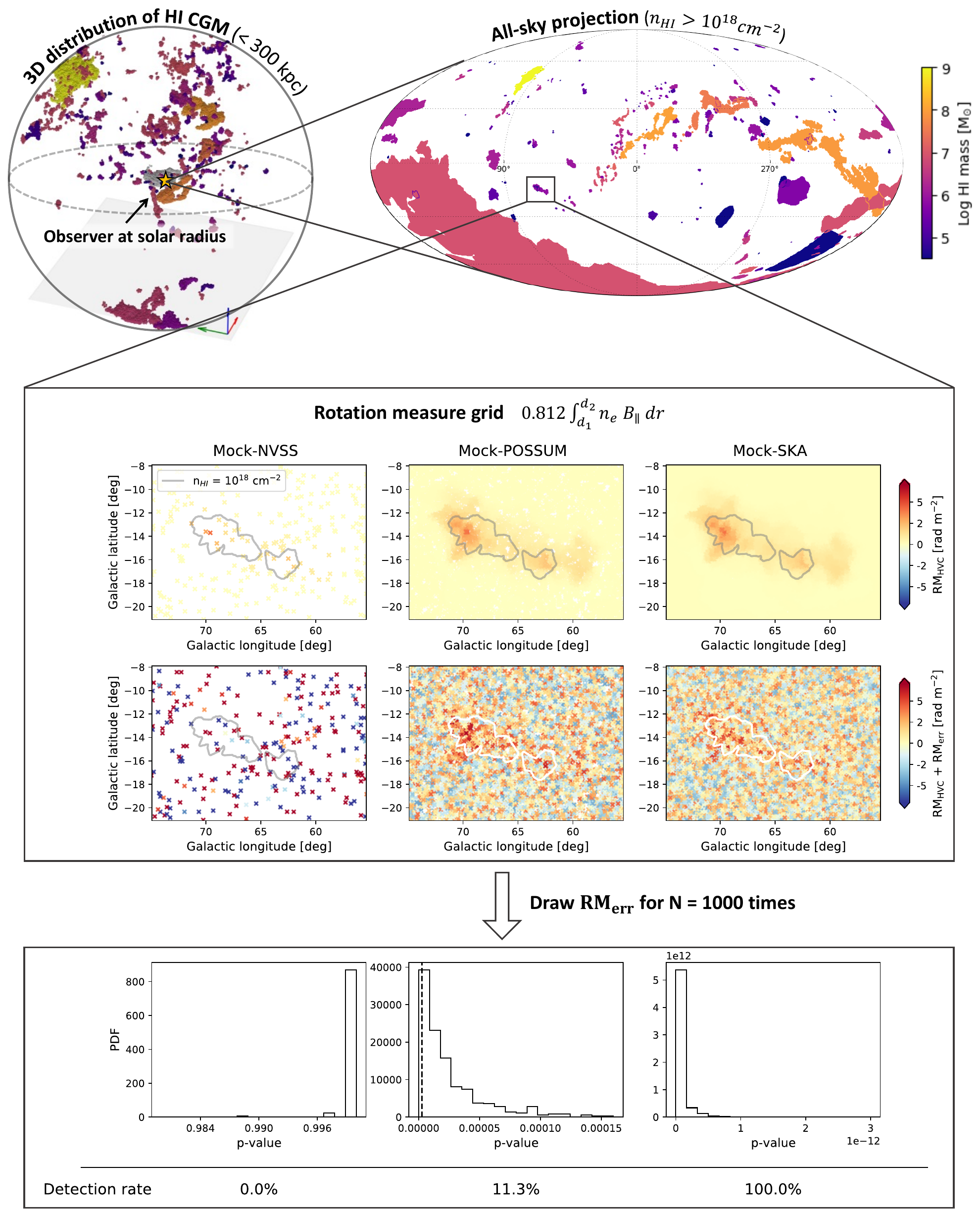}
    \caption{
    An illustration of how detection rates of individual magnetised circumgalactic clouds are calculated. 
    }  
    \label{fig:illust}
\end{figure*}

The top left panel of Fig. \ref{fig:illust} shows the 3D distribution of the identified HI circumgalactic clouds surrounding one of the sample galaxies (halo id: 82). Individual clouds are coloured by their HI mass in log-scale and the galactic disk at the centre of the sphere is shown in grey colour. The top right panel shows the all-sky distribution of the clouds in the Mollweide projection viewed by a mock observer inside the galaxy. The colour of each cloud is the same as the panel on the left and the boundary of the clouds shown in this panel is where the HI column density is higher than $n_{\rm HI}>10^{-18}\,\rm cm^{-2}$. 

Various physical properties of cold CGM clouds in TNG50 simulations are explored by \citet{Ramesh_2023b}. Although we do not adopt the same cloud identification criteria as their work, we expect the properties of our clouds to be overall similar to what is presented in their paper. 

\section{Detection statistics for magnetised HI clouds}\label{sec:result1}

\subsection{Obstacles for the detection}\label{sec:obstacle}

There are several obstacles that make observations of magnetised circumgalactic clouds using RM grids challenging. We start by introducing some of them.
\begin{enumerate}
    \item \underline{\textit{Limited polarised source density of RM grids}}\\
    A necessary condition for the detection of magnetised clouds is to have a statistically meaningful number of polarised sources in the background of the cloud of interest. 
    The source density of RM grids is decided based on the sensitivity of the polarimetric observation. 
    Based on deep observations of faint extragalactic polarised sources, \citet{Rudnick_2014} estimate the number distribution of the polarised sources follows
    \begin{equation}\label{eq:source_density}
    N(>p)\sim 45*(p/30\,\rm \mu Jy)^{-0.6}
    \end{equation}
    at $1.4\,\rm GHz$, where $p$ is the detection limit of the polarised intensity and $N(>p)$ is the number of polarised sources above a certain intensity $p$ per square degree. The RM source catalogue of the NRAO VLA Sky Survey (NVSS, \citealt{Condon_1998}; \citealt{Taylor_2009}) has $\approx 1\,\rm source/deg^{2}$. Thanks to its wide sky coverage (82\%), NVSS has been a major contributor to the discovery of candidates for magnetised HVCs (\citealt{McClure-Griffiths_2010}; \citealt{Hill_2013}). However, any cloud complexes of an angular size less than a few square degrees are left out of systematic searches using NVSS as the RM grid density is not sufficient to ensure enough polarised sources overlap with the clouds to draw a statistically firm conclusion.
    As observation sensitivity improves, upcoming polarimetric surveys are expected to provide greatly enhanced polarised source densities. 
    
    \item \underline{\textit{RM measurement error}}\\
    The precision of RM measurements derived using the RM synthesis technique (\citealt{Burn_1966}; \citealt{Brentjens_2005}) is defined as 
    \begin{equation}\label{eq:rm_err}
        \left|RM_{\rm err}\right| = \frac{\delta \phi}{2 (S/N)},
    \end{equation}
    where $S/N$ is the polarised signal-to-noise ratio conventionally set to be $S/N\gsim6$ for reliable RM measurements (\citealt{Macquart_2012})
    and $\delta \phi$ is the resolution of the Faraday spectra, i.e., the full-width half maximum of the RM spread function, which can be estimated as
    \begin{equation}\label{eq:rmsf}
        \delta \phi \approx 2\sqrt{3}/ (\lambda_{\rm max}^{2}-\lambda_{\rm min}^{2}),
    \end{equation}
    where $\lambda_{\rm max}$ and $\lambda_{\rm min}$ are the upper and the lower limit of the wavelength coverage of the polarimetric observation.
    For signals from magnetised clouds to be confirmed with adequate statistical significance, the ensemble average of the RM produced by the clouds needs to be overall sufficiently larger than the error distribution.
    
    \item \underline{\textit{Complex physical structure of clouds}}\\
    MHD models of fast-moving clouds in a weakly magnetised medium commonly show magnetic field lines draped around the clouds (e.g., \citealt{Konz_2002}; \citealt{Dursi_2008}; \citealt{Gronnow_2018}). 
    \citet{Betti_2019} successfully demonstrate that the observed RM pattern across the Smith Cloud resembles what is expected when projecting a simple draped magnetic field configuration into a 2D plane.
    However, it should be noted that many MHD simulations mentioned above assume a spherically symmetric cloud with a uniform-density core, whereas observations clearly show clumpy structures in HI HVCs. 
    There are suggestions that the complex density structure of a cloud leads to complex magnetic field configuration as magnetic field lines can drape individual overdensities of the cloud (\citealt{Jung_2023}). 
    Such complex $n_{\rm e}$ and $B_{\parallel}$ structures along lines-of-sights through a cloud can cancel out much of the RM excess produced along the sightlines.
    Clouds in cosmological simulations self-consistently form and evolve. Therefore, such internal cancellations of RM signals are inherently taken into account in our detectability estimates in this study as long as the resolution of the simulations allows.
    
    \item \underline{\textit{Confusion from the Galactic foreground}}\\
    As indicated in equation \ref{eq:RM}, any electron overdensities or enhanced magnetic fields between a source and an observer contribute to the observed RM of a polarised source. As we are surrounded by the Milky Way ISM, any sightlines towards extragalactic polarised sources inevitably suffer from possibilities for confusion from the Galactic foreground.
    Our ability to identify magnetised circumgalactic clouds strongly depends on how well we subtract the Faraday rotation at foreground Faraday screens in the region of interest.
    Simple models of the Milky Way foreground RM structure, such as a 2D surface fit to sources off the cloud-of-interest (e.g., \citealt{Hill_2013} in the Smith Cloud region), are sometimes sufficient to take care of the Milky Way foreground.
    However, such a simple approximation does not hold when the foreground Faraday screens are more complex.
    One example of such a complex field is the Magellanic Leading Arm region, which happens to have a supernova remnant in the foreground with a strikingly similar angular size to the Magellanic Leading Arm (\citealt{Jung_2021}). In this case, it is impossible to draw definitive conclusions about the magnetic field properties of the Magellanic Leading Arm from RM measurements from extragalactic sources alone. 
    The determination of the true Galactic Faraday rotating foreground is highly topical and often advances through investigations of diverse observational tracers across a range of wavelengths (e.g., \citealt{Jung_2021}). 
    Smoothed all-sky RM maps (e.g., \citealt{Oppermann_2012, Hutschenreuter_2022}) are potentially useful for approximating large-scale coherent patterns (above the order of a few degrees) in all-sky RM. Because of their large angular scale, these patterns most likely trace global ISM characteristics. There are ongoing attempts to incorporate tracers of local ionized ISM distribution, such as using bremsstrahlung and $H\alpha$ emission to disentangle information about the Galactic ISM magnetic fields from smoothed RM distributions of extragalactic sources (\citealt{Hutschenreuter_2020, Hutschenreuter_2023}).
    Meanwhile, fluctuations in the RM at smaller scales (\citealt{Haverkorn_2008}; \citealt{Stil_2011}) are harder to constrain due to the stochastic nature of the turbulent magnetised ISM. 
    For simplicity, in this paper, we assume the contribution of the Galactic ISM to RM measurements has been taken care of in a complete manner.

    \item \underline{\textit{RM variations of various origins}}\\
    Similarly to points (ii) and (iv) above, the RM excess generated by magnetised HVCs needs to be large enough to stand out from other sources generating observed RM variations (e.g., medium directly associated with sources themselves and the IGM) in order to hold the statistical significance of detection.
    According to \citet{Schnitzeler_2010}, the observed standard deviation in RM independent of the Galactic latitude is $\approx 6.2\,\rm rad\,m^{-2}$. Although this scatter has been often referred to as an \textit{extragalactic contribution}, a part of the scatter inevitably comes from the Milky Way CGM, which is also independent of the Galactic latitude. 
    We will continue the discussion on this topic in Section \ref{sec:result2}.
    
\end{enumerate}

\subsection{Detection rates of clouds in the simulations}\label{sec:detection}

\begin{table*}
\centering
\caption{A summary of three synthetic Faraday samplings used in this study: Mock-NVSS, Mock-POSSUM, and Mock-SKA. The first column is the sampling density per square degree and the second column is the mean of the $\left|RM_{\rm err}\right|$ distribution inserted into the samples. The third column is the detection rates for magnetised circumgalactic clouds identified in the simulations. In columns 4 and 5, we show the detection rates when limiting the sample to nearby clouds ($<30\,\rm kpc$) and massive clouds (HI mass $>10^{6}\,\rm M_{\odot}$).}\label{tab:sampling}
\begin{tabular}{cccccc}
\hline
& Sampling density [$\rm deg^{-2}$]& $\left<\left|RM_{\rm err}\right|\right>$ [$\rm rad\,m^{-2}$] &  \begin{tabular}[c]{@{}c@{}}Detection rate for \\all clouds\end{tabular}  &  \begin{tabular}[c]{@{}c@{}}Detection rate for\\clouds within $30\,\rm kpc$\end{tabular} & \begin{tabular}[c]{@{}c@{}}Detection rate for\\clouds above $10^{6}\,\rm M_{\odot}$\end{tabular}\\\hline\hline
Mock-NVSS & 1 & 12.9 & 2.9\%& 12.0\%&11.7\%\\\hline
Mock-POSSUM & 25 & 2.1 & 29.3\%& 71.2\%&55.4\%\\\hline
Mock-SKA & 60 & 1.9 & 41.5\%& 83.0\%&68.7\%\\
\hline
\end{tabular}
\end{table*}

In this section, we evaluate whether individual clouds identified in the simulations are detectable with the given precision and sensitivity of synthetic polarimetry observations, directly addressing points (i) and (ii) above.
To do so, we construct synthetic RM grids around the clouds with varying source densities (i.e., the number density of sightlines sampled by a mock observer inside simulated galaxies). 
Since cold circumgalactic clouds self-consistently form and evolve in the cosmological simulations, our analysis inherently takes into account the possible complexity of the magnetic field and density structures around the clouds (i.e., point iii above) as far as the resolution of the simulations allows.

\begin{figure}
    \centering
    \includegraphics[width=\columnwidth]{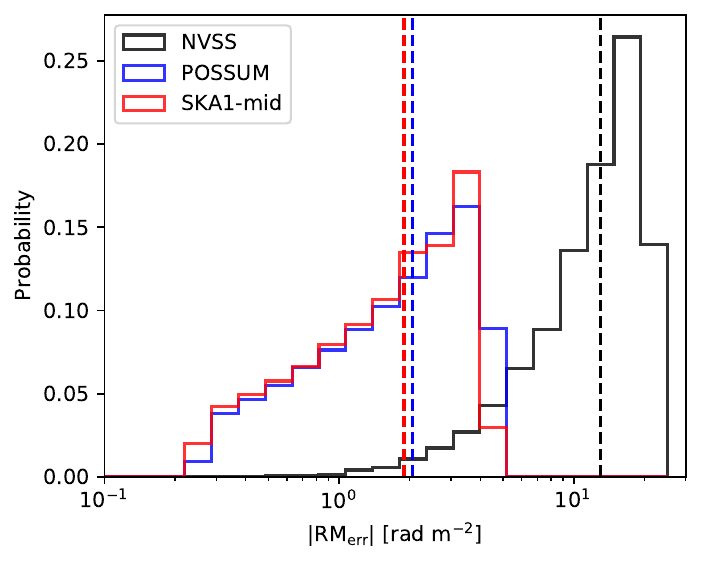}
    \caption{
    The histograms of $\left|RM_{\rm err}\right|$ for the NVSS (black), POSSUM (blue), and the SKA1-mid survey (red).
    The dashed lines show the mean of the distributions.
    The $RM_{\rm err}$ of the NVSS is the $1\sigma$ error of the observed sources multiplied by a factor of 1.22 (\citealt{Stil_2011}). For POSSUM and SKA1-mid survey, we construct expected $RM_{\rm err}$ distributions based on equations \ref{eq:source_density}, \ref{eq:rm_err}, and \ref{eq:rmsf}. See text for details.
    }  
    \label{fig:rm_err}
\end{figure}

Parameters for the synthetic samplings are chosen to emulate known specifications of current and upcoming polarimetric surveys; namely the NVSS, The Polarisation Sky Survey of the Universe's Magnetism (POSSUM) using the Australian Square Kilometre Array Pathfinder (ASKAP), and a future polarisation survey using the Square Kilometre Array (SKA1-mid). Below are brief descriptions of the surveys relevant to our synthetic sampling.
\begin{itemize}
    \item The NVSS RM catalogue (\citealt{Taylor_2009}) has on average one RM measurement per square degree.
    The black line in Fig. \ref{fig:rm_err} shows the distribution of $1\sigma$ error given by the catalogue multiplied by a factor of 1.22 (see \citealt{Stil_2011} for a reason for the scaling). The vertical dashed line of the same colour shows the mean value at $12.9\,\rm rad\,m^{-2}$.
    
    \item POSSUM is one of the major ongoing surveys of ASKAP (\citealt{Gaensler_2010}). The sensitivity ($\approx18\,\rm \mu Jy\,beam^{-1}$) and the bandwidth ($800-1088\,\rm MHz$; band 1) of the observations promise the average polarised source density of $\approx 25\,\rm deg^{-2}$ (based on equation \ref{eq:source_density}, see also \citealt{Anderson_2021}) or even higher\footnote{Note that \citet{Rudnick_2014} count sources observed at $1.4\,\rm GHz$. At the slightly lower frequency range of POSSUM ($0.9\,\rm GHz$), radio sources are in general slightly brighter and therefore POSSUM is likely to achieve an even higher source density.}.
    The expected width of the RM spread function is $54\,\rm rad\,m^{-2}$ (from equation \ref{eq:rmsf}). In Fig. \ref{fig:rm_err}, we show the distribution of expected $\left|RM_{\rm err}\right|$ of POSSUM sources and its mean in blue lines. 
    Descriptions of how we calculate the expected $RM_{\rm err}$ distribution will be presented shortly in the following paragraphs.

    \item As the Square Kilometre Array (SKA) is on the way, there are ongoing discussions on requirements and expectations for an optimal polarimetric survey in the SKA era. In this paper, we borrow the specifications of an RM grid survey present in \citet{Heald_2020}
    utilizing SKA1 mid-frequency band 2 ($950-1760\,\rm MHz$). Despite the slightly higher frequency range compared to POSSUM, the broader bandwidth will provide a narrower RM spread function, therefore, a slightly better RM accuracy. See the red line in Fig. \ref{fig:rm_err} for the expected distribution of $\left|RM_{\rm err}\right|$.
    The polarised source density of $\approx 60\,\rm deg^{-2}$ is expected given the suggested sensitivity of $4\,\rm \mu Jy\,beam^{-1}$.
\end{itemize}

The first two columns of Table \ref{tab:sampling} summarize the choice of the sampling density and the mean value of the $\left|RM_{\rm err}\right|$ distribution for each of our synthetic RM sampling (hereafter, Mock-NVSS, Mock-POSSUM, and Mock-SKA). We will explain the detection rate presented in columns 3, 4, and 5 shortly.

The central panels of Fig. \ref{fig:illust} show the three synthetic RM grids sampled on one of the HI circumgalactic clouds in the simulations. For each cloud identified in the simulations, we calculate $RM_{\rm HVC}$ along randomly selected sightlines by integrating $n_{\rm e} B_{\rm \parallel}$ as per equation \ref{eq:RM} (upper central panels in Fig. \ref{fig:illust}). The number of $RM_{\rm HVC}$ samples is decided based on the sampling densities.
We clarify that $RM_{\rm HVC}$ we refer to in this paper is the net Faraday rotation within a localized region enclosing a cloud. 
We effectively disentangle possible overlap between multiple clouds along a sightline by restricting the range of integration to $[d_{\rm min}-\Delta d, d_{\rm max}+\Delta d]$, where $d_{\rm min}$ and $d_{\rm max}$ are the minimum and the maximum span of a cloud in the distance domain and $\Delta d = 100\,\rm pc$ has been added as an additional buffer to the integration range.

Then, we incorporate the measurement errors of RM observations ($RM_{\rm err}$) to the pure $RM_{\rm HVC}$ as shown in the lower panels of the box in the middle of Fig. \ref{fig:illust}. 
For Mock-NVSS samplings, we randomly draw $RM_{\rm err}$ values from the observed error distribution of the NVSS catalogue and multiply by 1.22 (\citealt{Stil_2011}).
For Mock-POSSUM and Mock-SKA, we construct the expected $RM_{\rm err}$ distribution of each dataset based on the following procedures:
first, we utilize the polarised source count distribution observed by \citet[see equation \ref{eq:source_density} above]{Rudnick_2014} to obtain the polarised signal-to-noise ($S/N$) distribution of observable radio sources ($S/N\gsim6$). Then we plug in the $S/N$ values to equation \ref{eq:rm_err} and get the $\left|RM_{\rm err}\right|$ distribution for Mock-POSSUM and Mock-SKA separately. 
For simplicity, we assume that the polarised source count does not change significantly between the frequency range that \citet{Rudnick_2014} explored and what will be covered by POSSUM and SKA1-mid survey.

We define the ``detection'' of a magnetised cloud when the distribution of $RM_{\rm HVC}+RM_{\rm err}$ is statistically different from the $RM_{\rm err}$ distribution. We use the two-sample Kolmogorov-Smirnov (KS) test to quantify the difference between the two distributions.
The detection rate of a cloud is obtained by drawing different sets of $RM_{\rm err}$ from the $RM_{\rm err}$ distribution a large number of times ($N=10^{3}$).
For each draw, we calculate the p-value of the KS test, i.e., the degree of the difference between the $RM_{\rm err}$ distribution and the $RM_{\rm HVC}+RM_{\rm err}$ distribution.
The detection rate is defined as the probability of the two distributions being different with larger than $3 \sigma$ confidence.
\begin{equation}
    \text{Detection rate} \equiv  \frac{\text{N(p-value}<2.7\times10^{-6})}{10^{3}}.
\end{equation}
Note that we are performing multiple hypothesis testing and the Bonferroni correction is adopted to control the threshold for the global significance level. In this case, the p-values of individual hypothesis tests must be less than $2.7\times10^{-6}$ ($=0.0027/N$) to ensure the $3\sigma$ significance level. 
The histograms in the bottom box of Fig. \ref{fig:illust} show the distribution of p-values for each sampling experiment for the example cloud. The detectability of this cloud for example is $0\%$ for the Mock-NVSS, $11.3\%$ for the Mock-POSSUM, and $100\%$ for the Mock-SKA sampling.

In Table \ref{tab:sampling}, we provide the total detection rates of all clouds identified in the simulations (column 3), clouds within the distance of $30\,\rm kpc$ (column 4), and clouds above the HI mass of $10^{6}\,\rm M_{\odot}$, i.e., around the mass of the Smith Cloud or higher (column 5)
for each sampling.
Indeed, the detectability of magnetised clouds increases significantly from 2.9\% (mock-NVSS) to 29.3\% (mock-POSSUM) and 41.5\% (mock-SKA) with the improved source density (from $1\,\rm deg^{-2}$ to $25\,\rm deg^{-2}$ and $60\,\rm deg^{-2}$) and the characteristic RM precision ($12.9\,\rm rad\,m^{-2}$ to $2.1\,\rm rad\,m^{-2}$ and $1.9\,\rm rad\,m^{-2}$).
Comparing columns 3, 4, and 5, we find that the detection rate strongly depends on the distance to the clouds as well as the cloud mass. Magnetised clouds closer to an observer and/or more massive have higher chances of detection at a given RM sampling specification.

\begin{figure*}
    \centering
    \includegraphics[width=\textwidth]{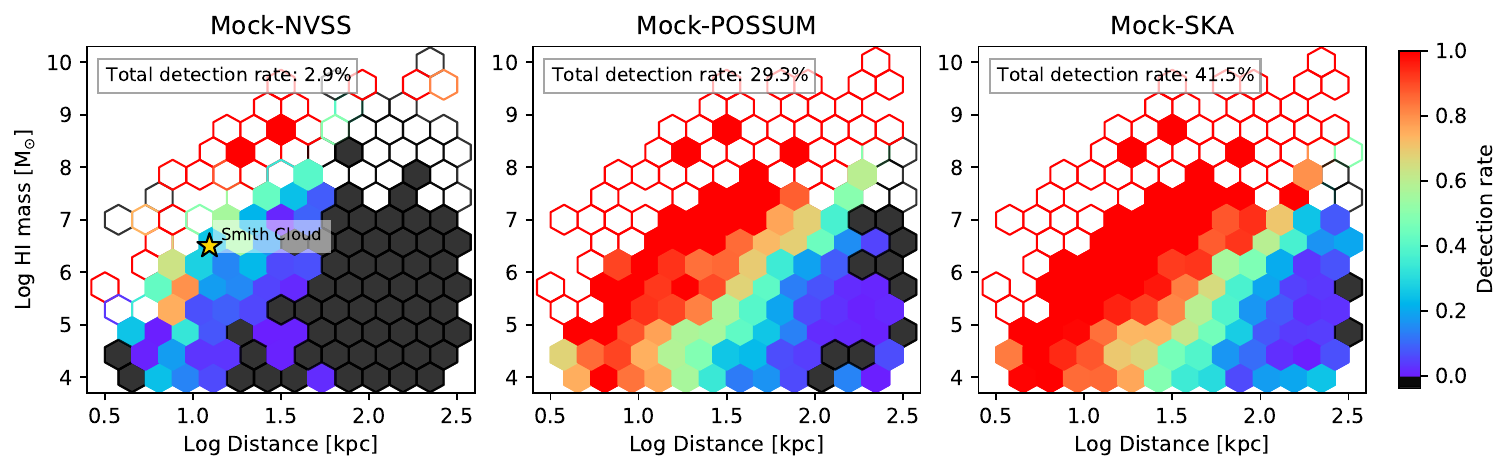}
    \caption{
    The detection rate of magnetised circumgalactic clouds as a function of the cloud HI mass and distance from an observer. Each panel shows results from different RM grid samplings (left: Mock-NVSS, middle: Mock-POSSUM, right: Mock-SKA). 
    The colour of each hexagon bin shows the mean detection rate of clouds within the distance and HI mass ranges. Cells with less than three clouds are shown as open hexagons.
    The detection rates are the highest among nearby massive clouds. The improved sampling capability (from Mock-NVSS to Mock-POSSUM and Mock-SKA) allows the detection of clouds with smaller masses and larger distances. The yellow star symbol in the left panel shows where the Smith Cloud is located in this parameter space.
    }  
    \label{fig:detect}
\end{figure*}

\begin{figure*}
    \centering
    \includegraphics[width=0.8\textwidth]{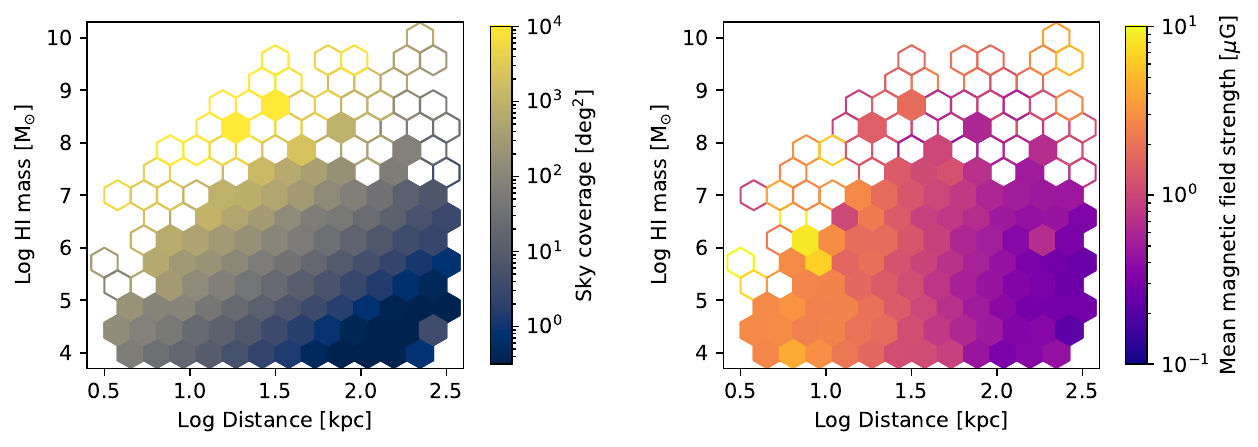}
    \caption{
    The sky coverage (left) and the mean magnetic field strength (right) of clouds are shown in the same format as Fig. \ref{fig:detect}. The closer and more massive the clouds are, the larger the solid angle. Also, clouds closer to the observer in general have larger mean magnetic field strength.
    }  
    \label{fig:detec_factors}
\end{figure*}

In Fig. \ref{fig:detect}, we present the detection rate of clouds as a function of the distance and the HI mass of the clouds. We take the distance between the observer and the clouds' centre of mass as a representative distance.
The cloud mass presented here is the HI mass, therefore, they are always smaller than the \textit{total} gas mass and depend on the HI fraction within a cloud.
Therefore, low-mass clouds in this figure (HI mass $\sim10^{4}\,\rm M_{\odot}$) do not conflict with the gas mass resolution of the simulations ($m_{\rm gas}=8.5\times10^{4}\,\rm M_{\odot}$, see Table \ref{tab:simul}). We discuss the resolution test in Appendix \ref{sec:resol}.
The colour of the hexagon bins shows the average detection rate of clouds in each bin.
Bins containing less than three clouds at the given parameter range are shown as open hexagons.

Results from the Mock-NVSS sampling (left panel) show that the detection rate of magnetised clouds is strictly limited to massive and closeby clouds. 
The yellow star in this panel is where the Smith Cloud, the only observationally identified magnetised HVC candidate unrelated to the Magellanic System, is located on this grid for reference (HI mass$\sim 10^{6.5}\,\rm M_{\odot}$ and distance $\sim12.4\,\rm kpc$, \citealt{Wakker_2008}; \citealt{Lockman_2008}).
The detectability estimated by the simulations at this parameter regime is fairly high ($\approx42\%$), indicating that the detection of magnetic fields associated with the Smith-Cloud-like population is not uncommon.

Both Mock-POSSUM and Mock-SKA sampling results present significantly increased detection rates at all mass and distance ranges, suggesting the detections are feasible even for distant and low-mass clouds.
We first focus on clouds with HI masses above $>10^{7}\,\rm M_{\odot}$. Although not many clouds are located at this mass range (most of the hexagon bins are unfilled, i.e., enclose fewer than three clouds), they are almost always detectable with mock-POSSUM and mock-SKA samplings. Clouds in this regime are mostly satellite galaxies or extended outer disk structures of the central galaxy. 

At the lower mass range (HI mass $<10^{7}\,\rm M_{\odot}$), we find a strong mass and distance dependency of the detection rate.
This trend stems from both the observational and physical nature of the clouds.
In the left panel of Fig. \ref{fig:detec_factors}, we show the mean sky coverage of clouds in the same HI mass -- distance plane. Simply reflecting the inverse-square law of the solid angle, the larger the distance to a cloud the smaller it appears in the sky projection. Large angular size is favourable for the detection of magnetised clouds as it means that a cloud is sampled with a large number of background sources at a fixed polarised source density. 
On the other hand, the right panel of Fig. \ref{fig:detec_factors} shows the mean magnetic field strength of the clouds. 
The stronger magnetic field should increase the RM contribution of the cloud, making the ensemble average of on-cloud RMs higher.
We find a clear trend that clouds closer to an observer have stronger magnetic fields compared to the ones at large distances.
A similar result is reported by \citet{Ramesh_2023b} where the authors show cold clouds in the inner halo are dominated by magnetic pressure over thermal pressure in comparison to the clouds in the outer halo.
We speculate this trend is a combined result of (i) the strength of the ambient halo magnetic fields being stronger at the inner halo and (ii) the presence of clouds originating from the strongly magnetised galactic ISM environment in the disk-halo interface.

In summary, we learn the following from our synthetic RM sampling experiment on circumgalactic HI clouds in TNG50:
\begin{enumerate}
    \item With specifications of currently existing polarimetry surveys (e.g., NVSS), it is not unexpected that the detection of magnetised clouds has been only a handful and limited to clouds that are nearby (the Smith Cloud, \citealt{Hill_2013}) or associated with the Magellanic System (the Large Magellanic Cloud; \citealt{Gaensler_2005}, the Small Magellanic Cloud; \citealt{Mao_2008}; \citealt{Livingston_2022}, and the Magellanic Bridge; \citealt{Kaczmarek_2017}).
    \item Polarimetric surveys conducted with upcoming radio telescopes (e.g., ASKAP and SKA) will provide improved polarised source density and RM measurement accuracy that can significantly increase the overall detection rate of magnetised clouds.
    \item At a given RM grid sampling specification, the detection rate is higher for clouds that are more massive and/or closer to the observer. This is not only because of their larger sky coverage but also because they have stronger magnetic fields.
\end{enumerate}

\subsection{Higher-order statistical tracers of magnetised clouds}

\begin{figure*}
    \centering
    \includegraphics[width=0.95\textwidth]{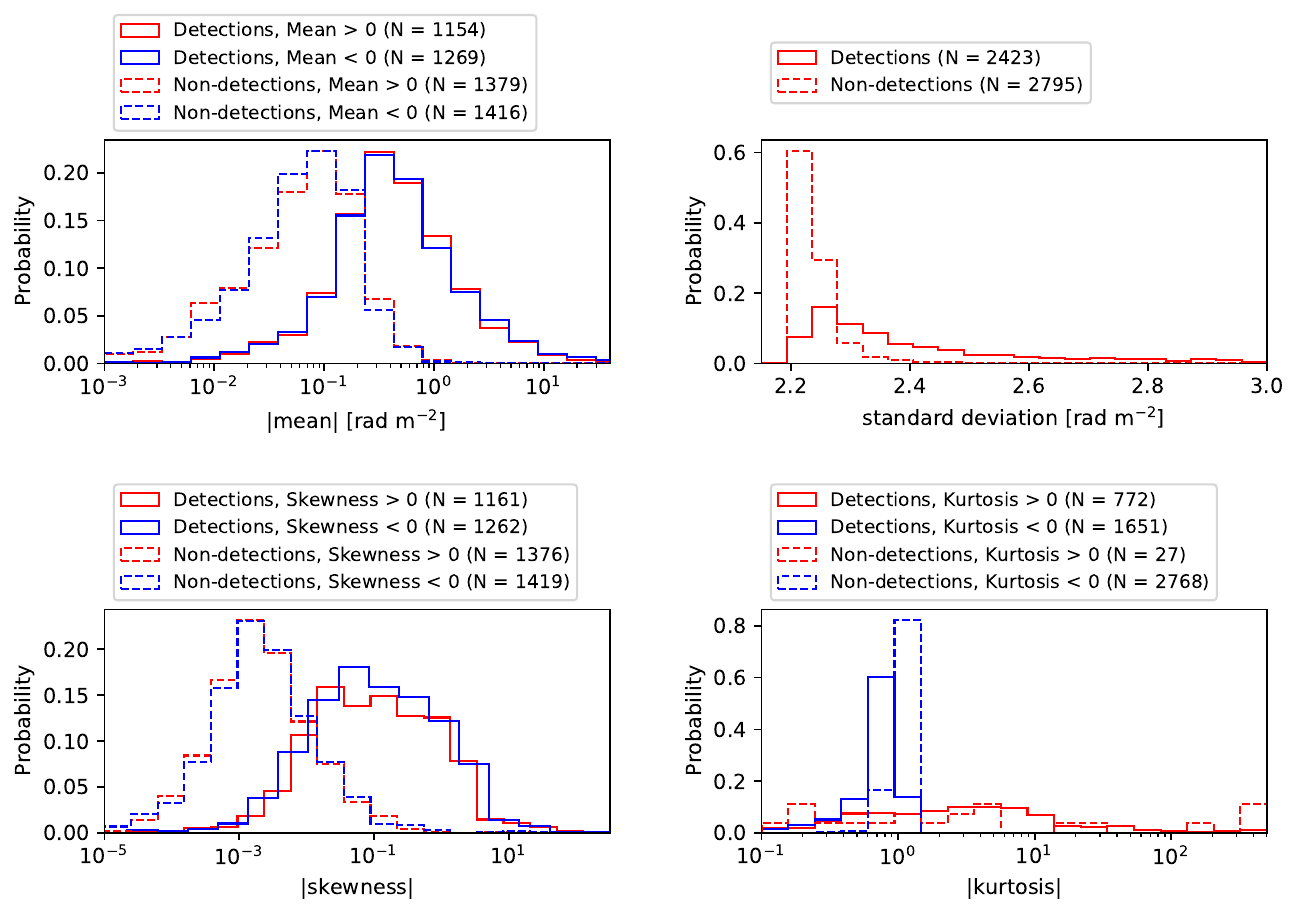}
    \caption{
    Each panel shows the mean (top left), standard deviation (top right), skewness (bottom left), and kurtosis (bottom right) of the RM distribution sampled around each cloud following Mock-SKA specifications. 
    We divide our cloud sample into four categories based on the detection rate (solid line if the detection rate is greater than zero and dashed line if it is zero) and the sign of the statistic displayed in the panel (red if it is greater than zero and blue if less than zero). We find a significant difference between detections and non-detections for the mean, standard deviation, and skewness.
    }  
    \label{fig:higher_stats}
\end{figure*}

Identifying the RM ``excess'' associated with cloud distribution in the sky
has been a widely used method to search for magnetised clouds using RM grids.
The excessive RM, i.e., a larger mean |RM|, is almost certainly a good tracer of overall enhancement of magnetic field strength in and around clouds (higher $B_{\rm \parallel}$ in equation \ref{eq:RM}). 
However, considering the RM is an observable parameter integrated along a sightline, the presence of magnetised clouds may not always produce a larger \textit{mean} RM in the region. 
For example, a complex 3D magnetic field geometry in a turbulent medium could cancel out locally enhanced RM when integrated along a sightline (see point iii in Section \ref{sec:obstacle} as well as Figure 1 of \citealt{Jaffe_2019}). 
In such cases, the imprint of a magnetised cloud would rather show up in higher-order statistics that trace fluctuations of RM at scales smaller than the scales probed by the RM grid.

In this section, we examine whether higher-order statistics of RM, namely standard deviation, skewness, and kurtosis, are useful tracers of magnetised clouds.
In Fig. \ref{fig:higher_stats}, we show histograms of each statistic calculated for each cloud in the simulations (top left: mean, top right: standard deviation, bottom left: skewness, bottom right: kurtosis). All parameters are measured within all-sky-projected rectangular regions that tightly enclose HI clouds.
In all panels, the colour of the histograms shows the signs of the statistics (red: positive, blue: negative). 
Furthermore, we separate the cloud sample into the detection and the non-detection groups in order to identify parameters that demonstrate a clear discrepancy between the two groups and consider them useful measures for finding magnetised clouds in RM grids.
The solid lines are for clouds with a non-zero detection rate according to the mock-SKA sampling and the dashed line histograms are for non-detections for comparison. 
Note that our definition of the detection depends on the type of sampling we use (mock-NVSS, mock-POSSUM, and mock-SKA). The change in the sampling type alters the number of clouds in the detection and the non-detection group, but we confirm that the distribution of each statistic does not depend strongly on the sampling type we use.

We start from the top left panel of Fig. \ref{fig:higher_stats}.
The absolute mean RM of the detection group (solid line) is overall higher than the non-detection group (dashed line), demonstrating that the RM excess in general serves as a useful tracer of magnetised clouds. 
Note that the measurement noise ($RM_{\rm err}$) distribution we insert to the pure $RM_{\rm HVC}$ distribution has the mean value of zero. Thus, when there is no systematic contribution of a cloud, the expected mean RM value is zero.
We do not see differences between the distributions of positive (red) and negative (blue) mean RM values. This indicates that there is no preferred line-of-sight direction of mean magnetic fields in the clouds.

The top right panel of Fig. \ref{fig:higher_stats} shows the distribution of RM standard deviation. The standard deviation is by definition always positive.
The distribution of the non-detection group (dashed line) peaks at $\sim 2.2\,\rm rad\,m^{-2}$, which simply reflects the standard deviation of the $RM_{\rm err}$ distribution that we insert to the $RM_{\rm HVC}$ of the mock-SKA sampling. 
The detection group (solid line) peaks at a slightly higher standard deviation and there is a long tail extended towards higher values. In this panel, we only show the distribution between 2.2 and $3\,\rm rad\,m^{-2}$, but it is worth noting that $25\%$ of the clouds in the detection group have the RM standard deviation higher than $3\,\rm rad\,m^{-2}$.

We present the skewness distribution in the bottom left panel of Fig. \ref{fig:higher_stats}.
The skewness parameterizes the asymmetry of the distribution. Zero skewness means the distribution is symmetric about the median.
A positively skewed distribution has a tail extended towards the higher value (positive RM in our case) and a negatively skewed distribution is extended towards the opposite side (negative RM). 
The distribution of RM around individual clouds is more skewed among the detection group (solid line) compared to the non-detection group (dashed line). 
We do not find differences in the distribution of absolute skewness between negatively (blue) and positively (red) skewed populations.
It is worth mentioning that among the clouds in the simulations, $85\%$ have the same signs of the mean and the skewness.

Finally, the bottom right panel of Fig. \ref{fig:higher_stats} shows the kurtosis distribution.
The kurtosis describes how extended the tails of a distribution are compared to the normal distribution. We adopt Fisher’s definition of excess kurtosis: if positive, the distribution approaches zero at both ends more slowly than a Gaussian and if negative, the tails fall faster than a Gaussian.
The majority of the clouds have negative kurtosis (blue) strongly peaked at the absolute value of $\approx1$, and the bias toward negative kurtosis is stronger for the non-detection group (dashed line, $99\%$) than for the detection group (solid line, $68\%$).
In comparison, the distribution of positive kurtosis values (red) is widely spread over several orders of magnitudes. We confirm that most clouds with large positive kurtosis are massive clouds (HI mass $\gsim 10^{7}\,\rm M_{\odot}$).
Overall, the difference in kurtosis between the detection and non-detection groups is subtle compared to other statistics inspected in this work.

From our analysis in this section, we conclude that magnetised circumgalactic clouds leave imprints on the mean, standard deviation, and skewness of the RM distribution, but not so much in kurtosis. 
The imprints in multiple RM statistics can be used as strong evidence of magnetised clouds in the absence of a strong detection in mean RM. 
One caveat is the correction of the foreground Galactic ISM which can potentially make a significant contribution to the observed RM statistics. As mentioned earlier in Section \ref{sec:obstacle}, characterising the Galactic foreground is beyond the scope of this paper and we have assumed that the foreground has been perfectly removed. 
While the foreground attributes of the observed Milky Way circumgalactic clouds must be taken care of on case-by-case bases (e.g., \citealt{Jung_2021}), constraints on the expected RM statistics of the foreground can be acquired from a better characterization of the turbulent properties of the ISM. 
This is a field of active ongoing investigations and has been defined as one of the main scientific goals of future radio polarimetric surveys (\citealt{Heald_2020}).

\section{Contribution of the CGM to the all-sky RM distribution}\label{sec:result2}

In this section, we now broaden our focus to the contribution of \textit{the entire CGM} to the all-sky RM distribution, not limited to regions enclosing HI clouds. In the following paragraphs, we provide further motivation to do so.

The majority of polarised point-like radio sources providing RM measurements are extragalactic objects.
Therefore, any observed RM measurements are a superposition of Faraday rotation taking place at any magneto-ionized structures between an observer and a source. For example:
\begin{equation}
\begin{aligned}
    RM_{\rm obs} = &RM_{\rm int} + RM_{\rm ex-gal} + RM_{\rm MW, CGM} + RM_{\rm MW, ISM}\\& + RM_{\rm err}.
\end{aligned}
\end{equation}
Each term of the above equation portrays Faraday rotation (i) intrinsic to the media within the vicinity of the source itself ($RM_{\rm int}$), (ii) at extragalactic structures like intervening galaxies and the large-scale structure ($RM_{\rm ex-gal}$), (iii) at the Milky Way CGM ($RM_{\rm MW, CGM}$), (iv) at the Milky Way ISM ($RM_{\rm MW, ISM}$), and (v) added to the signal due to the instrumental noise ($RM_{\rm err}$).
Similarly, the RM variance adds up assuming each RM component is statistically independent and follows the Gaussian distribution:
\begin{equation}\label{eq:rm_var}
\begin{aligned}
    \sigma_{\rm obs}^{2} = &\sigma_{\rm int}^{2} + \sigma_{\rm ex-gal}^{2} + \sigma_{\rm MW, CGM}^{2} + \sigma_{\rm MW, ISM}^{2} + \sigma_{\rm err}^{2}.
\end{aligned}
\end{equation}

Decomposing the contributions of individual components empirically from the observed RM distributions is challenging, but there have been attempts to do so.
\citet{Schnitzeler_2010} fit a Galactic latitude-dependent model to the RM dispersion measurements using the NVSS RM catalogue and separate latitude-dependent and latitude-independent components of the observed RM spread. 
The estimated $\sigma_{\rm RM}$ values are $\approx7.6\,\rm rad\,m^{-2}$ for the latitude-dependent component and $\approx6.2\,\rm rad\,m^{-2}$ for the latitude-independent component. 
The author refers to the former as a Galactic contribution and the latter as an extragalactic contribution to the RM dispersion with the caveat that the Milky Way can have latitude-independent components. \citet{Oppermann_2015} report the extragalactic RM dispersion of similar extent between $6.6-7.2\,\rm rad\,m^{-2}$.

Measuring the difference in RM between close pairs of radio sources provides independent estimations of the combined contribution of the intrinsic and extragalactic RM variations. 
This approach eliminates the Galactic contribution by assuming that two sources with small angular separation share the almost identical Galactic foreground Faraday screen.
\citet{Vernstrom_2019} use the NVSS RM catalogue and obtain the upper limit of the variation in RM.
One of the key findings of their work is that physically related pairs, e.g., two AGN lobes of one radio galaxy, have smaller $\Delta RM$ ($\approx 4.6\,\rm rad\,m^{-2}$) than random associations ($\approx14.9\,\rm rad\,m^{-2}$). This finding reinforces the explanation that the observed $\Delta RM$ originate from the extragalactic contribution. 
Similarly, \citet{OSullivan_2020} use the LOFAR Two-Metre Sky Survey (LoTSS) and find $\Delta RM \approx 1.4 - 1.8\,\rm rad\,m^{-2}$ between close radio pairs (see also \citealt{Pomakov_2022}). The authors attribute the small $\Delta RM$ to the low-frequency range ($144\,\rm MHz$) of the LoTSS data, where sources experiencing strong Faraday rotation suffer depolarisation. They argue such depolarisation effect in fact to some degree filters out radio sources influenced by unusually strong Faraday rotation intrinsic to the sources themselves, therefore, their measurements potentially better reflect low variance components of RM such as the cosmic web (\citealt{Carretti_2022}).
In regard to the RM dispersion intrinsic to radio sources ($\sigma_{\rm int}$), there are indications that $RM_{\rm int}$ can vary significantly source-by-source, over almost two orders of magnitude, and may systematically depend on source properties (\citealt{Rudnick_2019}).

From the numerical simulations' perspective, there have been efforts to estimate the contribution of the IGM and large-scale cosmic filaments to the observed RM. For example, \citet{Aramburo-Garcia_2021} use TNG100 to show that the integrated Faraday rotation within large-scale magnetised feedback bubbles in the intergalactic space can be as high as a few $\rm \mu G$.
\citet{Akahori_2011}  estimate the contribution of the extragalactic large-scale structures to the observed RM is $\sim 7-8\,\rm rad\,m^{-2}$ (see also \citealt{Akahori_2010}) which is comparable to the latitude-independent component of the observed RM spread estimated by \citet{Schnitzeler_2010}.
In their paper, the authors placed a mock observer within a Local Group-like environment which in part incorporates the contribution of the local halo environment. Yet, the primary focus of the simulations they utilize in their work is to reproduce realistic cosmic large-scale structures (the spatial resolution $=195h^{-1}\,\rm kpc$) and thus might not sufficiently reflect fluctuations in the local CGM which take place at a much smaller physical scale.
In any case, it is important to note that predictions from numerical simulations, including our own, have to be interpreted with caution as the exact extent of magnetic fields in cosmic structures and the estimated Faraday rotation may depend on how numerical simulations treat magnetic field seeding and amplification (\citealt{Vazza_2017}).

In this section, we will demonstrate that the Galactic CGM \textit{can} contribute significantly to the observed RM dispersion. We raise caution that the latitude-independent $\sigma_{\rm RM}$ estimated from observations is not necessarily dominated by signals from the extragalactic environment (e.g., $\sigma_{\rm int}$ and $\sigma_{\rm ex-gal}$ in equation \ref{eq:rm_var}), but potentially influenced by the spread in RM caused by the Galactic CGM ($\sigma_{\rm MW, CGM}$).

Although TNG50 is a cosmological suite that allows the study of cosmic large-scale structures and the Faraday rotation associated with them, we postpone this to future studies. As mentioned throughout this paper, we focus instead on the CGM of a galaxy surrounding an observer at redshift 0.
Accurate estimation of the contribution of large-scale structures 
on the observed RM requires consideration of various factors that are beyond the scope of this paper, such as (i) the evolution of magneto-ionic properties of the large-scale structure over a wide range of redshifts, (ii) the redshift distribution of polarised radio sources, and (iii) the scale factor dependency ($(1+z)^{-2}$) of the RM of polarised sources at cosmological distances (see, e.g., \citealt{Akahori_2011}).

\subsection{Measuring the variation in RM caused by the CGM}

\begin{figure*}
    \centering
    \includegraphics[width=\textwidth]{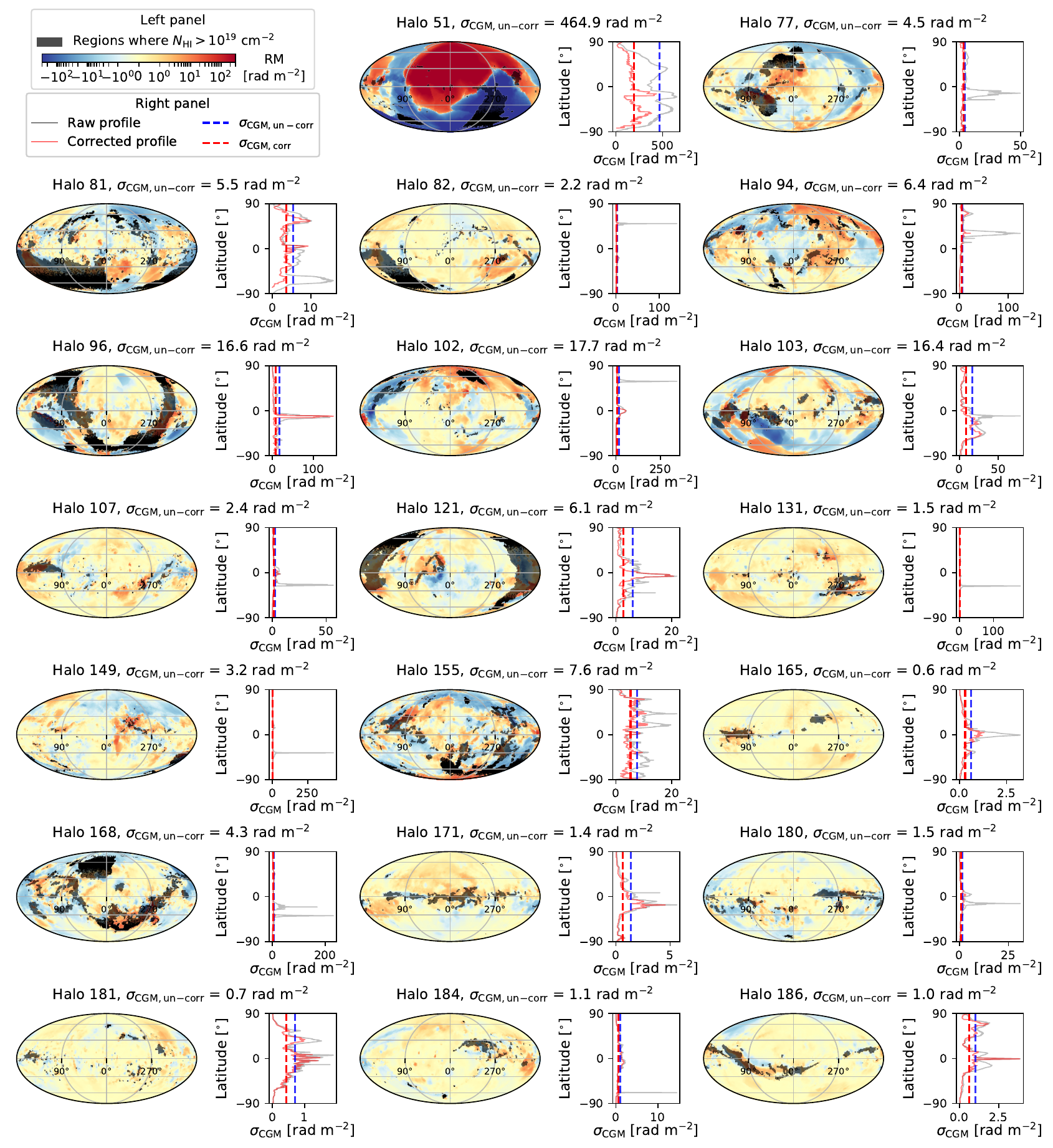}
    \caption{
    For each halo, the left panel is the Mollweide projection of the all-sky distribution of $RM_{\rm CGM}$. The gray shaded areas are where the HI column density is larger than $10^{19}\,\rm cm^{-2}$. We disregard RM measurements in these regions when calculating $\sigma_{\rm CGM}$. 
    The right panel shows $\sigma_{\rm CGM}$ as a function of the Galactic latitude. The grey line is the raw profile without any HI column density masking for reference. The red solid line is the profile after the correction for longitude dependencies as described in the text. 
    The two vertical dashed lines show $\sigma_{\rm CGM, un-corr}$ (blue) and $\sigma_{\rm CGM, corr}$ (red), respectively. See the text for the definitions of the two parameters.
    }  
    \label{fig:RM_allsky}
\end{figure*}

\begin{figure*}
    \centering
    \includegraphics[width=\textwidth]{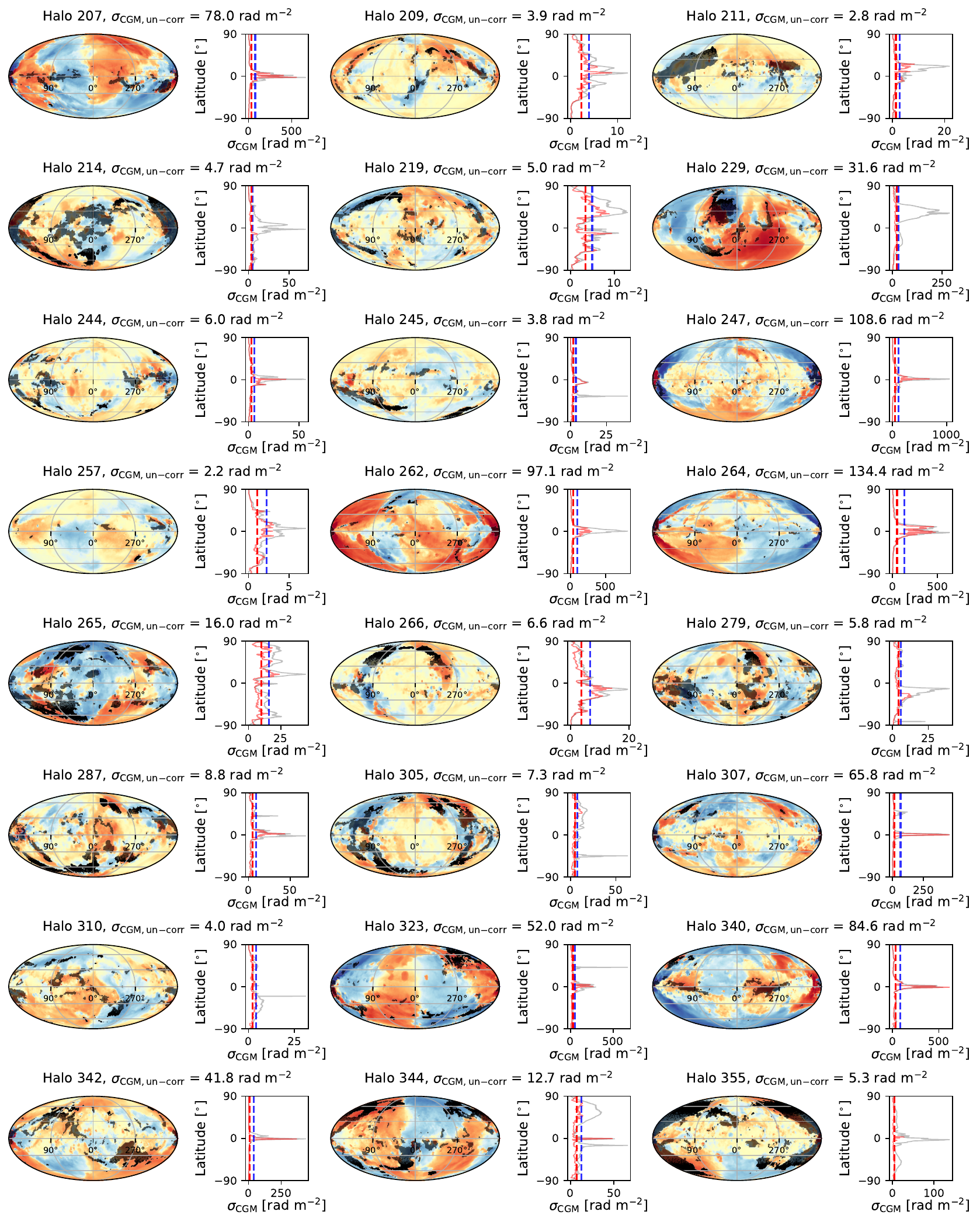}
    \caption{
    Continued from Fig. \ref{fig:RM_allsky}. 
    }  
    \label{fig:RM_allsky2}
\end{figure*}

\begin{figure*}
    \centering
    \includegraphics[width=\textwidth]{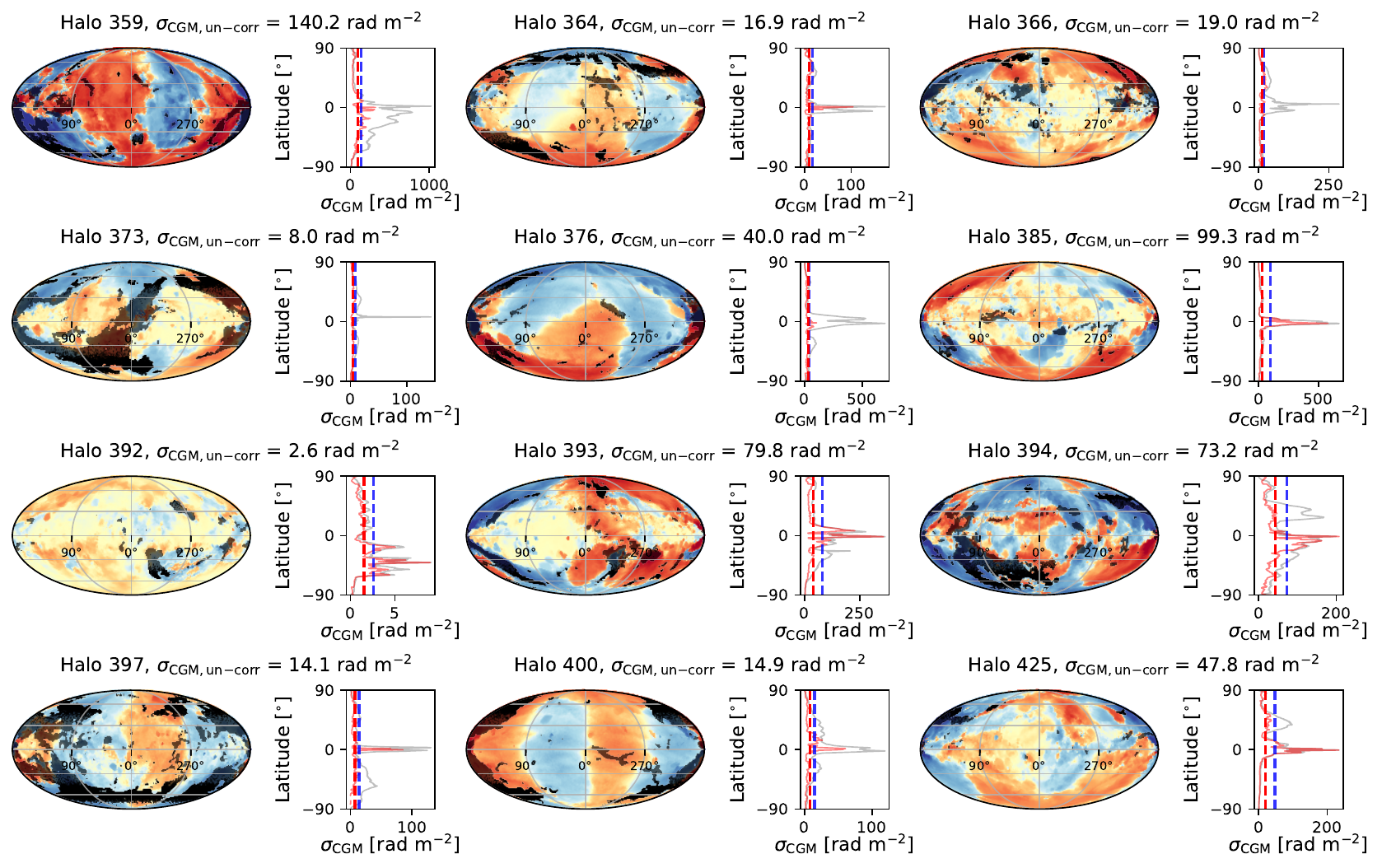}
    \caption{
    Continued from Fig. \ref{fig:RM_allsky2}.
    }  
    \label{fig:RM_allsky3}
\end{figure*}

\begin{figure}
    \centering
    \includegraphics[width=\columnwidth]{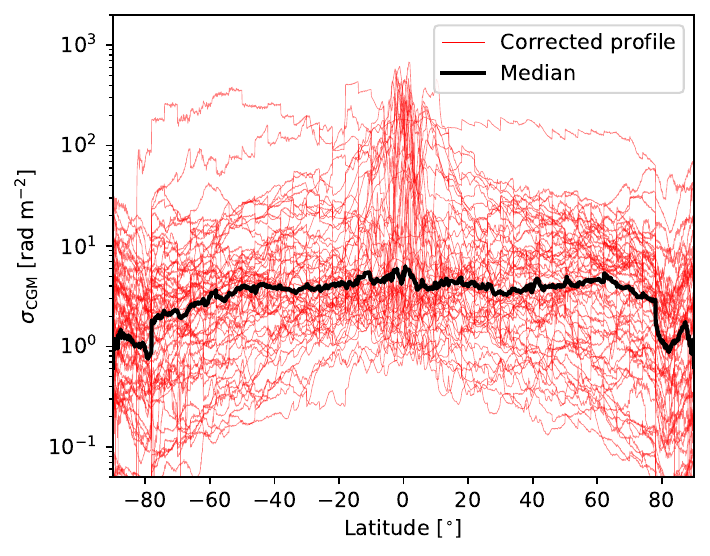}
    \caption{
    The longitude-corrected $\sigma_{\rm CGM}$ profiles of all the halos used in this work. The red lines are the same as the red lines in the right panels of Figs. \ref{fig:RM_allsky}, \ref{fig:RM_allsky2}, and \ref{fig:RM_allsky3}. The black line shows the median of all the profiles. 
    We confirm that there is no latitude dependency of $\sigma_{\rm CGM}$.
    }  
    \label{fig:sigrm_profile}
\end{figure}

In this paper, we calculate the spread in the all-sky RM distribution caused by the CGM ($\sigma_{\rm CGM}$) in two different ways, motivated by \citet{Schnitzeler_2010}. In both cases, we use the all-sky RM grid of the CGM uniformly sampled with the resolution of $(\Delta l, \Delta b) = (0.1^{\circ}, 0.1^{\circ})$ and $\Delta d=50\,\rm pc$. No measurement noise ($RM_{\rm err}$) is added to this sample as we are interested in the RM spread purely produced by the CGM in this analysis. 
\begin{itemize}
    \item The \textit{uncorrected} standard deviation (hereafter, $\sigma_{\rm CGM, un-corr}$) is a standard deviation of the all-sky $RM_{\rm CGM}$ distribution. Note that we have filtered out the galactic ISM contribution from the simulated galaxies (Section \ref{sec:ism_vs_cgm}). Therefore, we do not expect any latitude-dependent component of $\sigma_{\rm CGM}$ from our results. Also, we exclude any $RM_{\rm CGM}$ samples from the regions where the CGM HI column density is higher than $>10^{19}\,\rm cm^{-2}$ in order to avoid the contribution of obvious dense structures in the CGM localized to a certain region of the sky, such as satellite galaxies.
    \item The \textit{corrected} RM standard deviation (hereafter, $\sigma_{\rm CGM, corr}$) is the value we have calculated following the correction for longitude dependencies described in Section 3 of \citet{Schnitzeler_2010}. We provide a brief summary of the correction method here. The motivation behind presenting the corrected $\sigma_{\rm CGM}$ in this study is to ensure that we calculate the RM variance of the simulated sky as closely as possible to how it is calculated from observations. 
    We bin the RM grids along the galactic latitude: two polar cap regions above and below $b=\pm 78^{\circ}$ and 39 bands between $-78^{\circ}<b<78^{\circ}$ with the width of $\Delta b = 4^{\circ}$.
    Next, we further divide each cap/band along the galactic longitude and calculate the average RM within each cell. The cell size along the longitude is set to $\Delta l = 20^{\circ}$ for the polar caps and $\Delta l = 5^{\circ}/\cos(b)$ for the bands. The longitude-dependency of RM in each cap/band is determined using the cubic spline of the cell-averaged RM and then subtracted from the original RM grids. 
    From this longitude-corrected all-sky RM distribution, we calculate the standard deviation of the RM for $2^{\circ}$ bins in Galactic latitude and identify the representative $\sigma_{\rm CGM}$ that minimizes the $\chi^{2}$ of the 90 standard deviation measurements.
\end{itemize}

Figs. \ref{fig:RM_allsky}, \ref{fig:RM_allsky2}, and \ref{fig:RM_allsky3} consist of panels for each of the 56 Milky Way-like galaxies in our sample showing the all-sky $RM_{\rm CGM}$ distribution in the Mollweide projection (left) and $\sigma_{\rm CGM}$ as a function of the galactic latitude (right). 
As we have excluded the galactic ISM from our analysis to focus on the CGM, no obvious galactic disk is visible in the maps.
The grey-shaded regions in the left panels are where the HI column density of the CGM is higher than $>10^{19}\,\rm cm^{-2}$, i.e., areas excluded when calculating $\sigma_{\rm CGM, un-corr}$ and $\sigma_{\rm CGM, corr}$.

In the right panels, there are two solid lines showing the latitude profiles of the raw $\sigma_{\rm CGM}$ (grey; no high HI column density filtering and longitude correction applied) and the corrected $\sigma_{\rm CGM}$ (red), respectively. 
Two vertical dashed lines show $\sigma_{\rm CGM, un-corr}$ (blue) and $\sigma_{\rm CGM, corr}$ (red). 
A lot of the excess of $\sigma_{\rm CGM}$ in the raw $\sigma_{\rm CGM}$ profiles are associated with the locations of the high-column density regions of the sky, therefore, applying the HI column density filter effectively removes high $\sigma_{\rm CGM}$ peaks visible in the grey line, although some spikes are still present in the corrected profiles (red line). Such localized high RM variations in the corrected profiles are mitigated when taking the best-fitting value, $\sigma_{\rm CGM, corr}$, of all the latitudes (red vertical dashed line).
The corrected $\sigma_{\rm CGM}$ profiles are also presented in Fig. \ref{fig:sigrm_profile} where we confirm that there is no overall latitude-dependency of the $\sigma_{\rm CGM}$ since we have removed the galactic contribution.

\begin{figure*}
    \centering
    \includegraphics[width=0.9\textwidth]{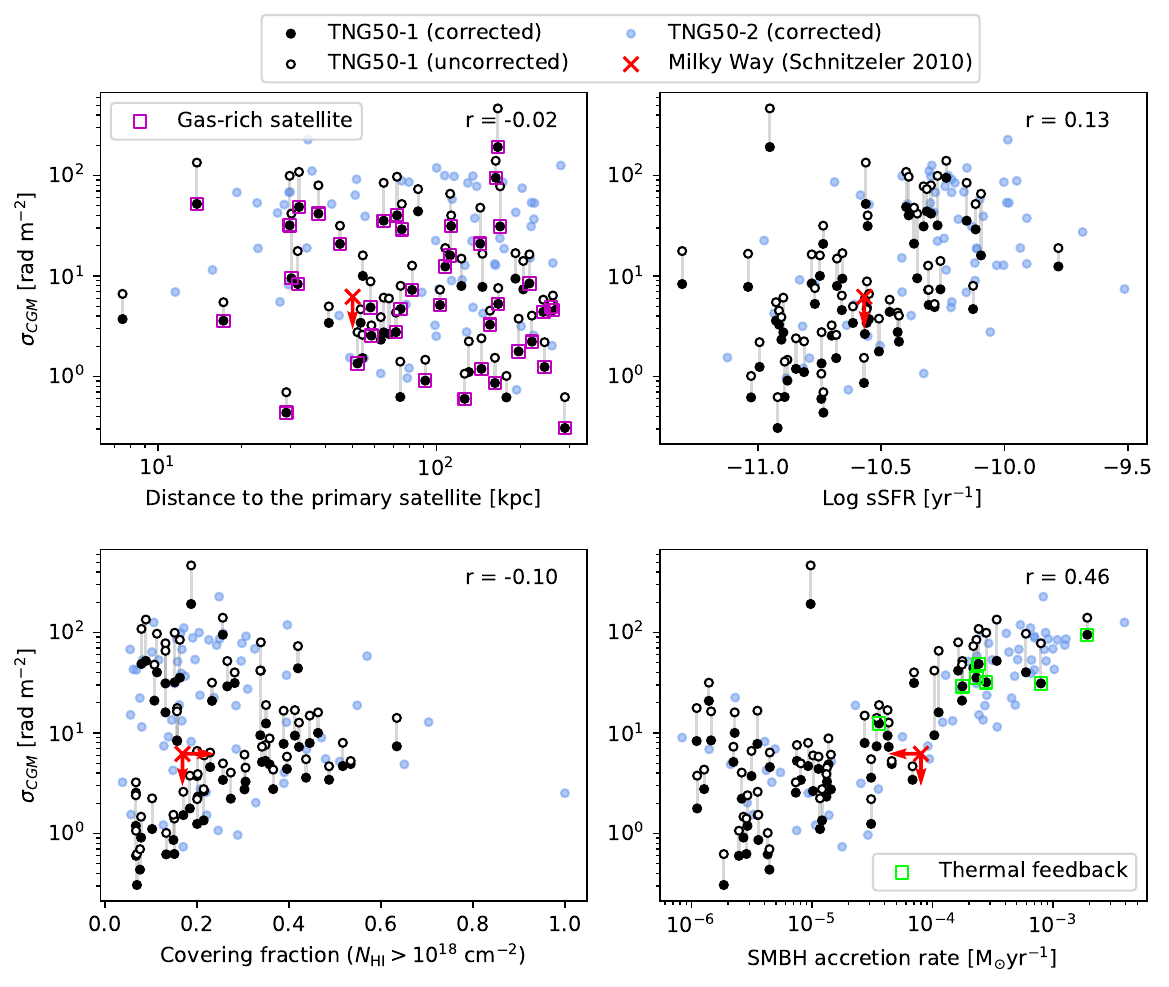}
    \caption{
    The RM variation ($\sigma_{\rm CGM}$) as a function of galaxy properties (top left: the distance to the most massive satellite galaxy, top right: the specific star formation rate, bottom left: the HI CGM covering fraction, and bottom right: the SMBH accretion rate).
    We mark the systems where the primary satellite is a gas-rich galaxy in the top left panel and the systems under thermal mode SMBH feedback in the bottom right panel to show that both populations are not distinguished from the overall distribution of the rest. 
    In all panels, the red cross symbol is the upper limit of $\sigma_{\rm RM}$ from the Milky Way observations (\citealt{Schnitzeler_2010}).
    The blue circles are the results from TNG50-2 simulations (lower resolution). We discuss the resolution effect in Appendix \ref{sec:resol}.
    The number in the upper right corner of each panel shows the Pearson correlation coefficient.
    }  
    \label{fig:sig_RM}
\end{figure*}

\subsection{Comparison with observations}

We now compare $\sigma_{\rm CGM}$ of the simulated galaxies with the observed value estimated by \citet[$\approx 6.2\,\rm rad\,m^{-2}$]{Schnitzeler_2010}. 
The observed RM spread we are comparing our results to is the latitude-independent component of the RM variance which possibly encompasses the combined contribution of $\sigma_{\rm int}$, $\sigma_{\rm ex-gal}$, and $\sigma_{\rm MW, CGM}$ terms in equation \ref{eq:rm_var}, whereas, from simulations, we are measuring the pure CGM contribution ($\sigma_{\rm CGM}$). Therefore, the measurement from the observations should be taken as an upper limit of the RM spread produced by the Milky Way CGM.

Even though we have attempted to select Milky Way-like galaxies in simulations using the criteria explained in Section \ref{sec:method}, $\sigma_{\rm CGM}$ varies by a lot among the sample. 
In order to understand the galaxy-by-galaxy variation of $\sigma_{\rm CGM}$, we investigate possible scaling relations between $\sigma_{\rm CGM}$ and various galaxy properties. 
In Fig. \ref{fig:sig_RM}, we present $\sigma_{\rm CGM}$ as a function of four parameters, all of which are well-constrained for both the Milky Way (red cross symbol) and the simulated galaxy sample. 
The filled black circles are $\sigma_{\rm CGM, corr}$ and open circles are $\sigma_{\rm CGM, un-corr}$ for comparison. The blue points are the measurements from TNG50-2, the lower-resolution simulation, that we will discuss in Appendix \ref{sec:resol} where we perform the resolution test.
Most notably, there is a large galaxy-by-galaxy variation in $\sigma_{\rm CGM}$, spanning almost two orders of magnitudes. The longitude correction for $\sigma_{\rm CGM}$ (filled circles compared to open circles) does reduce the $\sigma_{\rm CGM}$ of individual measurements, but it does not mitigate the spread among the galaxies. Below, we discuss what we find in each panel of Fig. \ref{fig:sig_RM} in more detail.

\begin{enumerate}
    \item \textit{Distance to the most massive satellite} (top left panel):\\
    The Milky Way is experiencing an ongoing accretion of the Magellanic System, which indeed appears to be leaving imprints in the observed RM sky, at the very least locally where the gas column density is high (\citealt{Gaensler_2005}; \citealt{Mao_2008}; \citealt{Kaczmarek_2017}; \citealt{Livingston_2022}). The distance to the Large Magellanic Clouds is $\approx50\,\rm kpc$ (\citealt{Pietrzynski_2019}). 
    In order to examine whether a close-by companion galaxy contributes to the spread in the all-sky RM distribution, we identify the most massive satellite galaxy within a halo and measure the distance to the galaxy. The stellar mass of the satellites varies between $10^{6}-10^{10}\,\rm M_{\odot}$.
    Not surprisingly, the RM measurements along sightlines through satellite galaxies are significantly higher than other regions of the sky due to the increased gas density and magnetic field strength.
    However, after masking out localized high HI column density on-satellite regions, we do not find a correlation between $\sigma_{\rm CGM}$ and the distance to the satellite.
    We further examine the distribution of galaxies whose primary satellite is a gas-rich galaxy (magenta square symbols), i.e., the total gas fraction is higher than 0.05, but there is no correlation between the satellites' gas fraction and $\sigma_{\rm CGM}$.
    
    \item{\textit{Specific star formation rate} (top right panel):\\
    The specific star formation rate (sSFR) is defined as the total star formation rate of a galaxy divided by the stellar mass.
    The observed sSFR of the Milky Way is $\approx 2.7\times10^{-11}\,\rm yr^{-1}$ (\citealt{Licquia_2015}).
    In simulations, we measure both the star formation rate and the stellar mass within the aperture of $30\,\rm kpc$ from the galactic centre. 
    As our definition of Milky Way-like halos takes the star formation rate as one of the criteria, our sample does not span a wide range of sSFR.
    Even so, there is a weak positive scaling relation between $\sigma_{\rm CGM}$ and sSFR within the sSFR range covered by our galaxy sample, though with a large scatter (the Pearson correlation coefficient $r = 0.13$). We consider this as an indication of a link between the sSFR and magnetic properties of the CGM. 
    This interpretation is in line with \citet{vandeVoort_2021} where the authors show that a galaxy and its outflows are coupled to the magnetic field in the CGM and vice versa.
    }
    
    \item{\textit{HI CGM covering fraction} (bottom left panel):\\
    In order to quantify the distribution of cold clouds in the CGM using an observable parameter, we measure the sky coverage of HI circumgalactic clouds with the column density limit of $>10^{18}\,\rm cm^{-2}$. 
    The covering fraction of the Milky Way HVCs is $\gsim20\%$, calculated from the HI HVC map of \citet{Westmeier_2018}.
    At this range of the covering fraction, simulated galaxies span a very wide range of $\sigma_{\rm CGM}$ ranging almost two orders of magnitudes. The spread in $\sigma_{\rm CGM}$ decreases going towards higher HI covering fractions. 
    We find a positive correlation between $\sigma_{\rm CGM}$ of simulated galaxies and the sky coverage calculated with an extremely low HI column density limit, say, $>10^{14}\,\rm cm^{-2}$, however, we do not present the result here as there are no observations sensitive to detect such a low column density HI CGM.
    }

    \item \textit{The supermassive black hole accretion rate} (bottom right panel):\\
    \citet{Quataert_1999} estimate 
    the upper limit of the gas accretion rate of Sagittarius A*, the supermassive black hole (SMBH) of the Milky Way, is $8\times10^{-5}\,\rm M_{\odot} yr^{-1}$. 
    There are studies demonstrating that, at least in the TNG50 suite, the SMBH feedback directly affects the gas composition and flow in the CGM of Milky Way-like galaxies (e.g., \citealt{Pillepich_2021}; \citealt{Ramesh_2023a}).
    We find $\sigma_{\rm CGM}$ overall moderately scales with the SMBH accretion rate (the Pearson correlation coefficient $r = 0.46$), though the scatter is large at lower accretion rates ($<10^{-5}\,\rm M_{\odot}yr^{-1}$).
    The TNG50 suite models the blackhole feedback in two modes depending on the SMBH mass and the accretion rate (\citealt{Weinberger_2017}): kinetic mode and thermal mode. We show galaxies under thermal mode feedback with green-coloured square symbols but do not find any appreciable differences in trends between the two groups. 
\end{enumerate}

We conclude that there is no single parameter that alone can explain the wide range of $\sigma_{\rm CGM}$ thus the all-sky RM fluctuations arise as a result of diverse processes related to the evolution of the CGM.
In all cases, the estimate from the Milky Way observations (red cross symbol) is located well within the scatter of the simulated galaxies. 
It is important to make it clear that we do not attempt to estimate the exact contribution of the CGM to the observed Milky Way RM variance. 
Instead, our result is a demonstration that in some galaxies similar to the Milky Way, Faraday rotations occurring within the CGM alone can produce RM dispersion similar to or even higher than observational estimates of all latitude-independent contributions combined. \textit{Therefore, it is possible that the RM dispersion that is intrinsic to radio sources themselves or coming from the IGM is smaller than previously considered.} 
Ongoing investigations to independently constrain intrinsic and extragalactic RM variations using current and future polarimetric observations will help disentangle this complication. 

\section{Summary}\label{sec:summary}

In this paper, we have explored the synthetic Faraday rotation sky of Milky Way-like galaxies in TNG50. 
We specifically focus on the Faraday rotation at the CGM of the galaxies and estimate its contribution to the observed RM distribution.
Here, we summarize the main points discussed in this paper. 

First, we evaluate the detectability of individual magnetised HI clouds in the CGM by quantifying whether the RM signal produced by the clouds is statistically distinguishable from the RM measurement error. 
In this synthetic RM sampling experiment, the main factors we consider are the polarised source density and the RM measurement accuracy. We construct three different RM grid samplings, namely mock-NVSS, mock-POSSUM, and mock-SKA by adopting the specifications of current and forthcoming polarimetric surveys. The currently available NVSS RM catalogue provides about one RM measurement per square degree with the precision of $\approx 12.9\,\rm rad\,m^{-2}$.
We expect significant improvements in both parameters in upcoming surveys, for example, POSSUM (25 sources per square degree with $\approx 2.1\,\rm rad\,m^{-2}$ precision) and SKA1-mid survey (60 sources per square degree with $\approx 1.9\,\rm rad\,m^{-2}$ precision).

The mock-NVSS sampling broadly reproduces the current status of the search for magnetised clouds using existing polarimetric observations, including NVSS; the detection is limited to nearby massive clouds (e.g. the Smith Cloud) and objects associated with infalling satellite galaxies (e.g. the Magellanic System). 
From our mock-POSSUM and mock-SKA sampling results, we predict a significant increase in the number of detections using upcoming surveys which will allow systematic studies of magnetised circumgalactic clouds. In all cases, the detection rate is particularly high for clouds that are close and massive. This trend results from both the increased angular sky coverage and the stronger magnetic fields associated with these clouds. 
Quantitatively, we expect an order of magnitude increase in the detection rate of magnetised clouds with POSSUM and SKA1-mid survey compared to NVSS. Simply scaling by this factor, we expect the number of observational confirmations of magnetised circumgalactic clouds to increase from about 4 (the Smith cloud and structures associated with the Magellanic System) to almost 40 with POSSUM and 50 with the SKA1-mid survey. 
Not only the upcoming surveys will find many magnetised clouds, but also their significantly improved polarised source density will open new opportunities to study the magnetic field structures of the clouds in great detail.

Although in this paper we primarily focus on synthesising NVSS, POSSUM, and SKA1-mid survey, our speculations about the improved power of upcoming surveys are certainly applicable to other polarimetric surveys that are already available or that will be coming very shortly. For example, The Rapid ASKAP Continuum Survey (RACS, \citealt{McConnell_2020}) has recently delivered the Spectra and Polarisation In Cutouts of Extragalactic Sources (SPICE-RACS, \citealt{Thomson_2023}) RM catalogue that has the polarised source density of $\approx4\,\rm deg^{-2}$ and the average RM accuracy of $1.6\,\rm rad\,m^{-2}$ above polarised $S/N>8$.
Also, it is worth mentioning that the SKA and its mid-band precursors, ASKAP and MeerKAT, are located in the southern hemisphere. Therefore, surveys ongoing or planned with telescopes in the northern hemisphere, e.g., 
the LOFAR Two-metre Sky Survey (LoTSS, \citealt{Shimwell_2019, O'Sullivan_2023}), 
the Apertif imaging survey (\citealt{Adams_2022}), 
and the Karl G. Jansky Very Large Array Sky Survey (VLASS, \citealt{Lacy_2020}), will have a significant contribution to accessing the CGM towards the northern sky. 

We further perform an evaluation of various statistics of the RM distribution as a tracer of magnetised clouds. Traditionally, the search for magnetised clouds using RM grids has mainly focused on identifying excessive RM, which usually refers to the enhanced magnitude of RM measurements among sightlines that point towards a cloud of interest. However, we suggest that future RM surveys with high source densities will be able to discover magnetised clouds using higher-order statistics, especially the standard deviation and skewness of the RM distribution as long as correction for the Milky Way ISM foreground can be done with reasonably high accuracy.

Finally, we study the degree of fluctuations in the all-sky RM distribution produced by the CGM. The observed spread in the RM distribution is an aggregation of any fluctuation introduced by Faraday rotating media between polarised sources and the observer, including the IGM, the Milky Way CGM, and the Milky Way ISM. The degree of importance of each individual component is difficult to estimate from the observations. 
By quantifying the RM variation produced solely by the CGM of the simulated Milky Way-like galaxies, we address the question of how much of the Galactic latitude-independent RM variance can be attributed to galactic/extragalactic components.

The simulated galaxies, even though we try to select Milky Way-like galaxies, show a wide spread of the all-sky RM standard deviation ranging two orders of magnitudes in $\rm rad\,m^{-2}$ unit. 
In view of the observationally demonstrated utility of Faraday rotation to measure magnetic fields in diverse extragalactic environments, we must count ourselves lucky that the Milky Way is not far more shrouded by a complicated CGM environment. 
We investigate the relationship between various global galaxy properties and the RM standard deviation, but we do not find a single galactic property that can explain the diversity. Instead, the RM variation in the CGM appears to be a combined result of various astrophysical processes governing the galaxy's evolution. 
One possible analysis we suggest for future studies is to trace how the observed RM fluctuations change over time in each galaxy and connect it to galactic processes.

The observed latitude-independent RM standard deviation reported by \citet{Schnitzeler_2010} falls well within the scatter in the distribution of the simulated galaxies. 
Considering that the observed value reflects the combined contribution of the Milky Way CGM and extragalactic structures, we cannot reject the possibility of the Milky Way CGM contributing significantly to the observed RM spread. In other words, the extragalactic/intrinsic RM variation may be smaller than what has been previously thought.

\section*{Data availability}

The TNG50 simulations are publicly available at \url{www.tng-project.org/data}. The key data products of this work (i.e., all-sky RM maps) have been produced with a library that is available upon request from Rüdiger Pakmor. The data directly related to this paper will be shared on reasonable request to the corresponding author S. Lyla Jung. 

\section*{Acknowledgements}

We thank the referee for their helpful report which improved the paper. 
We thank the IllustrisTNG collaboration for making the data publicly available.
Our analysis is performed using the Python programming language (Python Software Foundation, \url{https://www.python.org}). The following packages were used throughout the analysis: {\sc numpy} (\citealt{Harris_2020}),  {\sc SciPy} (\citealt{Virtanen_2020}), and {\sc matplotlib} (\citealt{Hunter_2007}). This research also made use of a publicly available {\sc pyfof} package (\url{https://pypi.org/project/pyfof/}).

NMMcG acknowledges funding from the Australian Research Council in the form of DP190101571 and FL210100039. SLJ, NMMcG, YKM, CLVE, and CSA acknowledge the Ngunnawal and Ngambri people as the traditional owners and ongoing custodians of the land on which the Research School of Astronomy \& Astrophysics is sited at Mt Stromlo.




\bibliographystyle{mnras}
\bibliography{mybib} 

\appendix

\section{Resolution test}\label{sec:resol}

\begin{figure}
    \centering
    \includegraphics[width=\columnwidth]{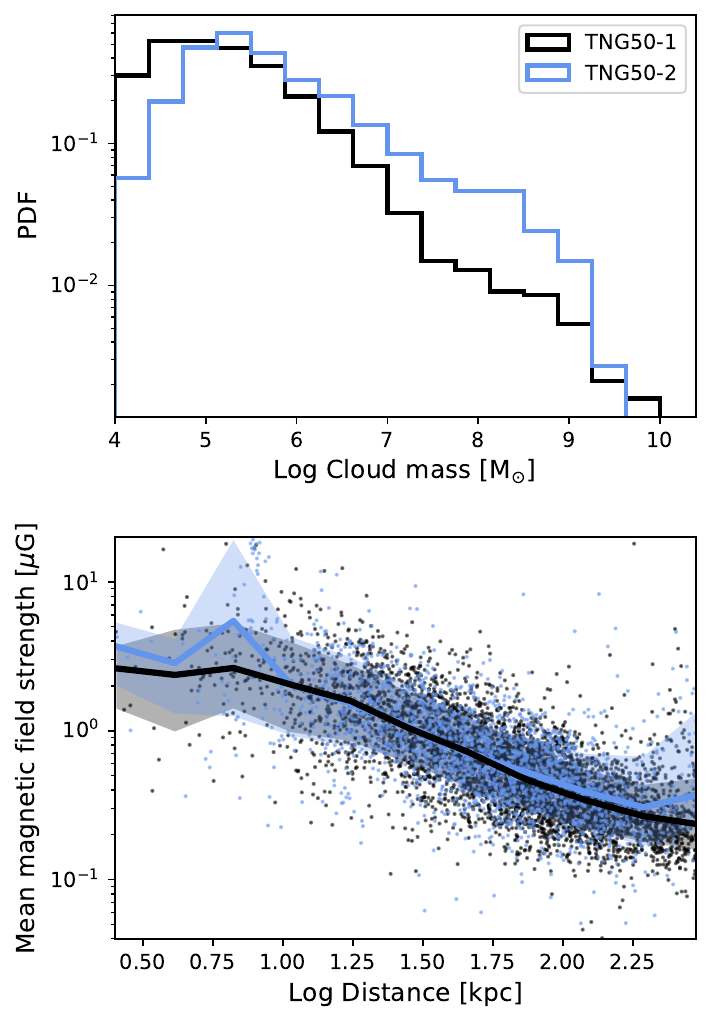}
    \caption{
    Top panel: the probability density distribution of the HI mass clouds identified in each simulation (black: TNG50-1, blue: TNG50-2).
    Bottom panel: the mean magnetic field strength within a cloud as a function of the distance to the cloud.
    The solid line shows the median value at a given distance and the shaded area encloses $68\%$ of the data points. There is no notable difference between the two simulations.
    }  
    \label{fig:dist_B_resol}
\end{figure}

The TNG50 suite consists of three simulations in identical settings and different resolutions. As for a resolution test of the results we present in this paper, we compare the fiducial simulation (TNG50-1) to the second lower-resolution simulation (TNG50-2). 
We explain the choice of parameters for each simulation in Table \ref{tab:simul} and the related text in Section \ref{sec:simul}. 

The last two columns of Table \ref{tab:simul} are the number of Milky Way-like galaxies and the number of circumgalactic clouds used for this study.
There are clearly fewer clouds identified in TNG50-2 (in total 2052) than in TNG50-1 (in total 5218), even though there are more Milky Way-like galaxies in TNG50-2 (in total 66) than in TNG50-1 (in total 56).
This is because lower-mass, smaller clouds are likely to suffer from the insufficient resolution of TNG50-2 (see also \citealt{Ramesh_2023b}).
Also, substructures of one cloud complex identified as multiple individual clouds in TNG50-1 are potentially merged into one large cloud in TNG50-2.
We show the HI mass distribution of clouds in the upper panel of Fig. \ref{fig:dist_B_resol}. Indeed, in TNG50-2 (blue), there is a higher fraction of massive clouds ($\sim 10^{7-9}\,\rm M_{\odot}$) in comparison to TNG50-1 (black). 
We also find a steep decline in the number of clouds in TNG50-2 at the low mass range ($\sim 10^{4-5}\,\rm M_{\odot}$) which is not as severe in TNG50-1.

Now, we compare a number of cloud properties directly related to the key conclusions of this paper.
Earlier in Section \ref{sec:detection}, we demonstrate that clouds closer to an observer in general have stronger mean magnetic field strengths (see Fig. \ref{fig:detec_factors}).
And that, along with an increased sky coverage, is one of the factors that makes nearby clouds more detectable in RM grids at a given RM sampling specification.
In the lower panel of Fig. \ref{fig:dist_B_resol}, we again present the mean magnetic field strength of clouds versus the distance, now comparing clouds in TNG50-1 (black) and TNG50-2 (blue).
Each data point corresponds to a single circumgalactic cloud identified in each simulation. The solid line is the median profile and the shaded region shows the $1\sigma$ scatter (encloses $68\%$ of the data points). 
We find that the clouds identified in TNG50-1 and those identified in TNG50-2 span the same range of mean magnetic field strengths and follow the same profile of decreasing magnetic field strength with increasing distance.


Earlier in Fig. \ref{fig:sig_RM}, we have presented all-sky $\sigma_{\rm RM}$ as a function of galaxy global properties. 
Focusing on comparing the results from TNG50-1 (black circles) and TNG50-2 (blue circles), we do not find a meaningful difference in the distribution of galaxies.
In TNG50-2, there are a larger number of galaxies with higher $\sigma_{\rm RM}$ ($\sim10^{2}\,\rm rad\,m^{-2}$), but this is because of their higher SMBH accretion rate ($\sim 10^{-3}\,\rm M_{\odot} yr^{-1}$, see bottom right panel) and is within the scaling relation also present in TNG50-1.

From the resolution test we present here, we conclude that the major results of this paper are not sensitive to the resolution.

\bsp	
\label{lastpage}
\end{document}